\documentclass[10pt,letterpaper]{article}
\usepackage[top=0.85in,left=2.75in,footskip=0.75in]{geometry}

\usepackage{amsmath,amssymb}

\usepackage{changepage}

\usepackage{textcomp,marvosym}

\usepackage{cite}

\usepackage{nameref,hyperref}
\usepackage{caption}

\usepackage[right]{lineno}

\usepackage[nopatch=eqnum]{microtype}
\DisableLigatures[f]{encoding = *, family = * }

\usepackage[table]{xcolor}

\usepackage{array}

\newcolumntype{+}{!{\vrule width 2pt}}

\newlength\savedwidth

\raggedright
\usepackage[aboveskip=1pt,labelfont=bf,labelsep=period,justification=raggedright,singlelinecheck=off]{caption}
\providecommand{\contcaption}[1]{\ContinuedFloat\caption[]{#1}} 

\makeatletter
\renewcommand{\@biblabel}[1]{\quad#1.}
\makeatother

\usepackage{lastpage,fancyhdr,graphicx}
\usepackage{epstopdf}
\fancyheadoffset[L]{2.25in}
\fancyfootoffset[L]{2.25in}
\usepackage{amsmath,amssymb}

\usepackage[table]{xcolor}
\usepackage{algorithm2e}
\RestyleAlgo{ruled}
\usepackage{import}

\newcommand{\argmax}[1]{\underset{#1}{\operatorname{argmax}}\;}

\newcommand{\be}{\begin{equation}}
\newcommand{\ee}{\end{equation}}
\newcommand{\bea}{\begin{eqnarray}}
\newcommand{\eea}{\end{eqnarray}}
\newcommand{\lpr}{\left(}
\newcommand{\rpr}{\right)}
\newcommand{\lbr}{\left[}
\newcommand{\rbr}{\right]}
\newcommand{\lcr}{\left\{}
\newcommand{\rcr}{\right\}}

\newcommand{\vb}{{\boldsymbol{b}}}

\newcommand{\vm}{{\boldsymbol{m}}}

\newcommand{\vo}{{\boldsymbol{o}}}

\newcommand{\vr}{{\boldsymbol{r}}}
\newcommand{\vs}{{\boldsymbol{s}}}

\newcommand{\vv}{{\boldsymbol{v}}}
\newcommand{\vw}{{\boldsymbol{w}}}
\newcommand{\vx}{{\boldsymbol{x}}}

\newcommand{\vJ}{{\boldsymbol{J}}}

\newcommand{\vM}{{\boldsymbol{M}}}

\newcommand{\vtheta}{{\boldsymbol{\theta}}}

\newcommand{\veta}{{\boldsymbol{\eta}}}

\newcommand{\vxi}{{\boldsymbol{\xi}}}

\DeclareMathOperator{\sigmoid}{\sigma}
\DeclareMathOperator{\ReLU}{ReLU}

\begin{document}
\vspace*{0.2in}

\begin{flushleft}
{\Large
\textbf{Inferring Inference} 
}
\newline
\\
Rajkumar Vasudeva Raju\textsuperscript{1,2},
Zhe Li\textsuperscript{2,3},
Scott Linderman\textsuperscript{4}, and
Xaq Pitkow\textsuperscript{1,2,5,6,7,8,*}
\\
\bigskip
\textbf{1} Department of ECE, Rice University, Houston, Texas, USA
\\
\textbf{2} Department of Neuroscience, Baylor College of Medicine, Houston, Texas, USA
\\
\textbf{3} Department of Ophthalmology, Stanford University, Stanford, California, USA
\\
\textbf{4} Department of Statistics, Stanford University, Stanford, California, USA
\\
\textbf{5} Neuroscience Institute, Carnegie Mellon University, Pittsburgh, Pennsylvania, USA
\\
\textbf{6} Department of Machine Learning, Carnegie Mellon University, Pittsburgh, Pennsylvania, USA
\\
\textbf{7} Center for Neuroscience and Artificial Intelligence, Baylor College of Medicine, Houston, Texas, USA
\\
\textbf{8} NSF AI Institute for Artificial and Natural Intelligence
\\
\bigskip

%
%





* xaq@cmu.edu

\end{flushleft}


%
\section*{Abstract}
Patterns of microcircuitry in the cerebral cortex suggest that the brain has an array of repeated elementary or ``canonical'' computational units. However, neural representations are distributed, so the relevant computations may only be related indirectly to single-neuron transformations. It thus remains an open challenge how to define canonical {\it distributed} computations. 
Here we integrate normative and algorithmic theories of neural computation to present a mathematical framework for inferring canonical distributed computations from large-scale neural activity patterns. At the normative level, we hypothesize that the brain has created a structured internal model of its environment, positing latent causes that explain its sensory inputs and using those sensory inputs to infer the states of the latent causes. At the algorithmic level, we propose that this inference process is a nonlinear message-passing algorithm on a graph-structured model of the world.
Given a time series of neural activity during a perceptual inference task, our analysis framework simultaneously finds $\lpr i \rpr$ the neural representation of the relevant latent variables, $\lpr ii \rpr$ interactions between these latent variables that define the brain's internal model of the world, and $\lpr iii \rpr$ message functions that specify the inference algorithm. Crucially, to be identifiable, this message-passing algorithm must use canonical nonlinear computations shared across the graph. 
With enough data, these targeted computational properties are then statistically distinguishable due to constraints from the symmetries inherent in any canonical computation, up to a joint global transformation of all interactions and the message-passing functions. As a concrete demonstration of this framework, we analyze artificial neural recordings generated by a simulated brain that implicitly implements an approximate inference algorithm on a probabilistic graphical model. Given external inputs and noisy neural activity from the simulated brain, we successfully recover the latent variables, their neural representation and dynamics, and canonical message functions that govern the dynamics. Finally, analysis of these models reveals features of experimental design required to successfully extract canonical computations from neural data. Overall, this framework provides a new tool for discovering interpretable structure in complex neural recordings.


%
\section*{Author summary}
Many features of the cerebral cortex suggest that the brain is built from repeated basic building blocks of computation. However, because information in the brain is spread across many neurons rather than handled by single cells, it has been difficult to identify what these basic computations actually are. This makes it challenging to connect detailed neural recordings to clear, interpretable computational principles.
In this work, we introduce a new framework for uncovering shared, fundamental computations directly from large-scale neural activity data. Our approach is based on the idea that the brain builds an internal model of the world, inferring hidden causes that explain sensory inputs. We propose that this process operates as a message-passing algorithm, where information is exchanged between parts of this internal model using a standard, repeated computation.
Using recordings of neural activity during perception, our method can simultaneously identify the hidden variables the brain represents, how those variables are related, and the rules governing how information flows between them. Tests on simulated neural data demonstrate that the method successfully recovers the underlying structure and computations. We also provide guidance about what types of experiments are best for revealing these canonical computations. Overall, this framework offers a new way to extract meaningful, interpretable structure from neural recordings.


\section*{Introduction}
We hypothesize that emergent computations in the brain are lawful and obey compressible rules: lawful, because dynamics are governed by a few canonical nonlinear operations, repeated in some form across many inputs or conditions; and compressible, because those operations can be summarized with far fewer parameters than needed to describe arbitrary dynamics. More specifically, we propose that these computations define a message-passing probabilistic inference algorithm. We also assume that this structure is hidden in the collective dynamics of many neurons. In this paper, we develop a framework for discovering this hidden message-passing algorithm within large-scale measurements of neural activity, and demonstrate a first application of this framework to a simulated brain to infer its underlying inference algorithm.

A {\it canonical} computation is a fundamental operation that repeats across brain regions and modalities to apply the same operations in a variety of contexts \cite{miller2016canonical,carandini2012normalization,AtomsofNeuralComputation}. Consider two levels at which these canonical computations can occur: (\textit{i}) the neural circuit/implementation level and (\textit{ii}) the algorithmic level \cite{marr1982vision}. Circuit level canonical computations include mechanisms like feedforward inhibition, divisive normalization, coincidence detection, gating information between different cortical areas, and working memory storage \cite{carandini2012normalization,AtomsofNeuralComputation}. A vanilla recurrent neural network is a simple example of this, where nonlinear activation functions act as circuit level canonical operations.  In contrast, here we propose that other canonical computations emerge strictly at the algorithmic level, hidden within the collective population dynamics of many neurons. These emergent canonical computations could be even more interpretable and perceptually relevant than local canonical mechanisms.

A prime example of such an algorithmic level canonical computation is {\it message-passing}. Message-passing reduces complex, often intractable global computations over many related variables into a sequence of simpler local operations. For example, Google's PageRank \cite{page1999pagerank} is a well-known message-passing algorithm that estimates the importance of web pages by distributing and recombining estimates of web page values along web links. Message-passing algorithms work by iteratively sending information, or messages, along edges in a graph relating those variables. Because message-passing algorithms use the same operations for all nodes and edges on a graph, they are an example of canonical computation at the algorithmic level.
Although message-passing can be used for any graph-structured computation, we are particularly interested in probabilistic inference on a graph-structured model of the world.

To discover such computations from neural recordings, our work integrates normative, algorithmic, and mechanistic theories \cite{marr1982vision} of recurrent nonlinear processing in the brain. We develop a structured statistical model for fitting neural data and revealing these distributed computational motifs, and demonstrate how to connect these theories to the mechanistic level. 

At the normative level, we hypothesize that the brain has an internal model of the world, positing latent variables that explain its sensory data. The Good Regulator Theorem states that every good regulator of a system must be a model of that system \cite{conant1970every}, and the No Free Lunch Theorems \cite{wolpert1997no} show that no algorithm can be superior to any other when averaged over all possible tasks or stimulus ensembles. Whether from extensive training, inherited innate knowledge, or a combination of the two \cite{zador2019critique}, one expects a sufficiently flexible intelligent system will successfully create an internal model that parallels the relevant structure of its natural environment, in particular by representing a set of sparsely interacting latent causes \cite{goyal2022inductive}. One natural manifestation of this general framework is a Bayesian brain \cite{helmholtz1925treatise,doya2007bayesian} that represents complex probability distributions using a formalism called probabilistic graphical models: these models use graphs to describe conditional independence relationships between latent variables \cite{koller2009probabilistic}, and algorithms can exploit these graphs for efficient inference.

At the algorithmic level, we propose that inference in the brain is implemented by nonlinear dynamics describing the flow of statistical information through a graph-structured internal model of the world \cite{raju2016inference,pitkow2017inference}. 
Specifically, we posit that these dynamics are structured as message-passing, which the brain uses to synthesize its sensory evidence and choose actions. 

At the mechanistic level, to connect this message-passing algorithm to data, we must specify how neural activity represents these latent variables and how computations are implemented through neural dynamics. Although single neurons are the physical substrate of computation, it has become clear that neural representations are highly distributed \cite{saxena2019towards,ebitz2021population,whiteway2019quest,yuste2015neuron}. Therefore, we argue that distributed processing is the appropriate level of abstraction for understanding this hypothesized message-passing algorithm.

There are two distinct senses in which a neural code could be distributed: information about a single variable could be distributed across many neurons (a population code \cite{georgopoulos1986neuronal}), and single neurons could contain information about many variables (mixed selectivity \cite{rigotti2013importance}).
Both properties are thought to coexist in the brain. This makes it harder to understand structured processing. It would be much easier to understand if the brain dedicated distinct populations purely to distinct variables \cite{sabour2017dynamic,dulberg2022modularity}. Although localist codes have long assumed such dedicated modular architectures \cite{zeki1974functional,felleman1991distributed}, ample evidence shows that task-relevant representations are not always localized \cite{saxena2019towards,musall2019single}. When structured computation does not directly parallel structured anatomy, new methods are needed to extract and describe these computational structures \cite{langdon2025latent}.

\begin{figure}[ht!]
	\centering
	\includegraphics[width=1\textwidth]{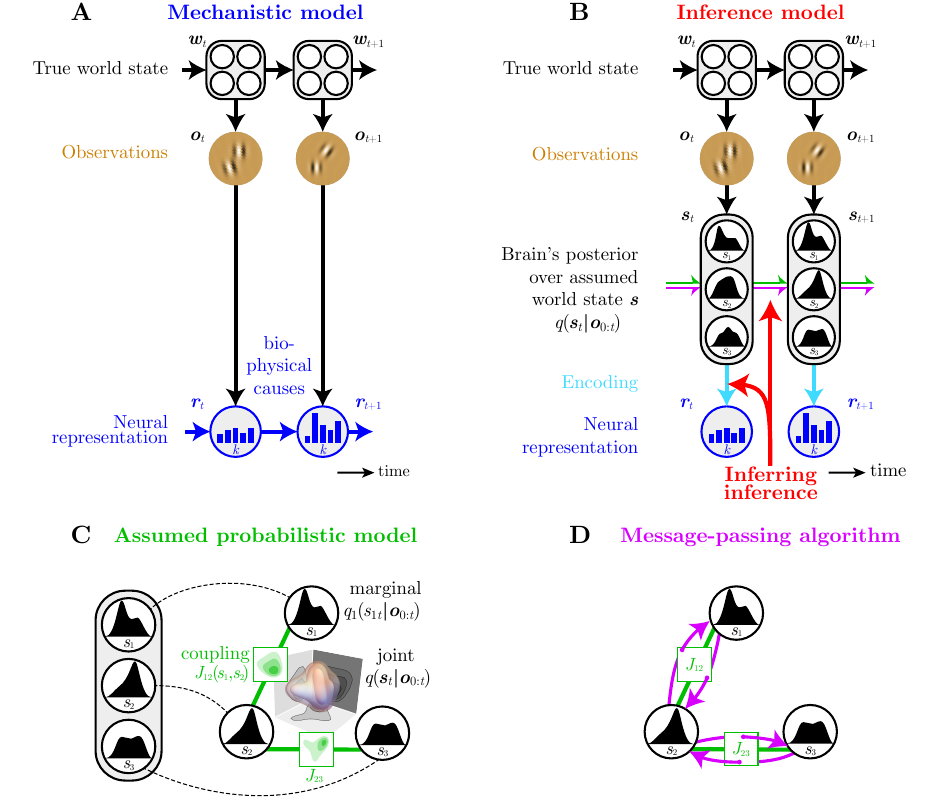}
	\caption{{\bf Schematic of inferring inference}. {\bf A}:  Mechanistic model has observations $\vo_t$ at time $t$ generated from an evolving world state $\vw_t$. These observations affect the intrinsic dynamics of neural responses $\vr_t$. {\bf B}: We impose a low-dimensional normative interpretation upon these mechanistic dynamics, where the reduced dimensions encode approximate posterior probabilities $q$ over assumed task-relevant variables $\vs_t$, which may differ from the true world state $\vw_t$. {\bf C}: The dynamics are structured according to a sparsely connected probabilistic graphical model, where signals propagate through task-relevant dimensions only along edges in the underlying graph. {\bf D}: This propagation is assumed to be lawful, following a canonical nonlinear message function $\mathcal{M}$ that is shared by all edges, but is modulated by coupling strengths that are specific to each edge.
	}
	\label{fig:Schematic}
\end{figure}

Our work provides a conceptual framework for interpreting how collective neural population dynamics are structured to perform behaviorally relevant computations via message-passing. This framework is illustrated in Fig~\ref{fig:Schematic}, which maps the relationship between the mechanistic model (A) and the algorithmic components (B--D). In the following sections, we introduce the mathematical formalism underlying these components.

\section*{Results}

\subsection*{Message-passing as a model for inference in the brain}

To discover the brain's computations, we assume that relevant canonical operations are shared across an entire unknown graph of unknown interacting variables. We aim to infer these latent states, their coupling, and how they evolve over time. Although our proposed framework is technically agnostic about the meaning of the neural responses, we are motivated by the hypothesis that the brain performs approximate Bayesian inference. We formalize this by mapping our three theoretical levels --- normative, algorithmic, and mechanistic --- into a unified mathematical framework. First, we describe the brain's internal model as a probabilistic graphical model \cite{koller2009probabilistic}. Second, we describe the canonical message-passing dynamics that performs inference on this model. And third, we establish how these latent computational dynamics could manifest in neural population activity. 

\subsubsection*{Normative level: probabilistic graphical model}
Aligned with the rich history of the Bayesian brain hypothesis, we assume that the brain uses a mental model of the world to  unconsciously perform probabilistic inference \cite{helmholtz1925treatise,doya2007bayesian}. We elaborate on this general idea by assuming that the brain structures its mental model as a probabilistic graphical model (PGM) \cite{lee2003hierarchical,koller2009probabilistic,vilares2011bayesian,haefner2016perceptual,pitkow2017inference,vertes2018flexible}.
PGMs elegantly specify structured relationships between variables that matter in tasks, using a graph to specify conditional dependencies.
We use PGMs to represent probability distributions $p(\vs)$ over a vector of latent variables $\vs$.  Crucially, we distinguish these assumed latent variables $\vs$, which constitute the brain's internal model, from the true world state $\vw$. While the true world state $\vw$ generates the sensory observations $\vo$, the brain must estimate its assumed variables $\vs$ conditioned on these observations, yielding the posterior distribution $p(\vs|\vo)$.

For simplicity, this paper focuses on pairwise undirected graphical models. We address generalizations to dynamic, directed, causal graphs with higher-order interactions in the discussion. A pairwise undirected PGM $\mathcal{G}$ uses nodes $\mathcal{V}$ and edges $\mathcal{E}$ to represent a probability distribution $p(\vs|\vo)$. Each node $i$ 
represents one variable $s_i$. These variables interact with each other along edges $\lpr i,j \rpr$ that indicate conditional dependencies. The joint distribution is described by the Boltzmann distribution $p(\vs|\vo)\propto e^{-E(\vs|\vo)}$ with a total energy $E(\vs|\vo)$ that decomposes into a sum of simpler terms, one for each node and edge. Critically, where there is no edge, the interaction energy contributed by that corresponding pair of disconnected variables is zero, and those variables are conditionally independent given all other variables.
This graph structure imposes restrictions that allow us to represent complex multivariate distributions in terms of a simpler structure.

\subsubsection*{Algorithmic level: message-passing by canonical computations}

Natural tasks usually involve only a subset of all variables, and often just a single variable $s_i$. Thus, an intelligent agent benefits from computing the marginal probabilities $p_i(s_i|\vo)$ of each variable to select optimal actions. Marginalization is therefore a central inference problem for the brain, serving as a concrete example of a canonical computational problem. However, in general, exact marginalization is intractable, requiring integration over all variables except the relevant one. Instead of performing this intractable computation by brute force, the brain may approximate these computations, leading to approximate posterior marginals $q_i(s_i|\vo)$ instead of the true ones.

Message-passing is a natural category of algorithms for approximate marginalization, because it can exploit the graph structure of the probabilistic graphical model to perform these important inference computations efficiently \cite{pearl1988probabilistic,wainwright2008graphical,yoon2018inference,zhang2020factor,fei2024higher}. Our motivating theory assumes that the brain follows this strategy, and we would like to uncover the computational dynamics leading to $q_i$. We assume that a state vector $\vx_i$ can parameterize the marginal distribution $q_i$, such that some decoding function $\tilde{q}$ could map these parameters onto the approximate posterior marginal: $\tilde{q}_i(s_i|\vx_i)\approx q_i(s_i|\vo)$ \cite{yoon2018inference,fei2024higher}.

To formalize these algorithmic dynamics, we adopt the general structure of a graph neural network (GNN) \cite{scarselli2009graph,kipf2018neural,battaglia2018relational} to describe the dynamics of a generic message-passing algorithm. 
Each node $i$ in the undirected graph $\mathcal{G}$ is associated with the vector-valued state $\vx_i$, and each edge $\lpr i, j \rpr$ with a vector-valued parameter $\vJ_{ij}$.
On every time step, each node sends a \textit{message} to each of its neighboring nodes. The message $\vm_{j\to i,t+1}$ from node $j$ to $i$ at time step $t+1$ depends on the state of the source and destination nodes and their edge parameter:
\begin{equation}
\vm_{j\to i, t+1} = \mathcal{M}\lpr \vx_{jt},\, \vx_{it},\, \vJ_{ij}\rpr
\label{eq:message_function}
\end{equation}
where $\mathcal{M}$ is the message function.
Next, an aggregation function $\mathcal{A}$ combines the incoming messages into a single message for the destination node:
\begin{equation}
\vM_{i \, t+1} = \mathcal{A}\lpr \lcr \vm_{j \to i,t+1} ~;~ j \in {\rm Nei}_i \rcr\rpr
\label{eq:aggregation_function}
\end{equation}
where ${\rm Nei}_i$ is the set of all neighbors of node $i$ in $\mathcal{G}$. Aggregation functions are typically permutation-invariant, {\it i.e.} they depend only on the set of incoming messages, and not on which node they come from (although these messages themselves do depend on the source nodes and edge parameters). Example aggregation functions include a simple sum, mean, or product. 
Finally, every node updates its state based on its previous state, any current external inputs $\vv_{it}$ to that node, and the aggregated message:
\begin{equation}
\vx_{i \, t+1} = \mathcal{U}\lpr \vx_{it}, \vv_{it}, \vM_{i\,t+1}\rpr
\label{eq:node_update_function}
\end{equation}
where $\mathcal{U}$ is the node update function, and $\vv_{it}$ is an external input at node $i$ at time $t$. Since the transition to the next state $\vx_{t+1}$ depends collectively on the past only through the current state $\vx_t$, these message-passing dynamics are Markovian.

An important property of these dynamics is that the message-update, aggregation, and node-update functions each have the same form at all locations on the graph. The specific messages vary with context, and the message function depends on edge parameters which may differ between edges, but all messages have the {\it same} dependence on edge parameters. This imposes a canonical organization at the algorithmic level, or what we could call an algorithmic symmetry. 

Many distinct inference algorithms --- such as mean-field inference \cite{opper2001advanced}, belief propagation \cite{pearl1988probabilistic}, expectation propagation \cite{minka2001expectation} --- all arise from different nonlinear transformations, characterized by the choice of functions in a message-passing algorithm. Even Gibbs sampling can be viewed as a message-passing algorithm with a stochastic update.
Canonical nonlinear functions in the biological brain might conceivably implement even smarter variants that are well-suited to the natural environment \cite{pitkow2017inference}. Our goal is to identify the canonical functions that define the brain's inference algorithms.

\subsubsection*{Mechanistic level: neural manifestations of latent dynamics}
At the mechanistic level, information processing is implemented by the collective behavior of neural populations. Any description of this processing should be consistent with established properties of neural codes, in which information is distributed across redundant neurons with mixed selectivity \cite{rigotti2013importance,moreno:2014information}. Thus, we posit that the brain's internal model of the world is not directly reflected in the activity or connections of individual neurons. Rather, it is \emph{implicit} in the mechanistic interactions between possibly overlapping neural populations \cite{semedo_cortical_2019,langdon2025latent}. It is therefore crucial to quantify how information is represented and transformed in a low-dimensional latent space embedded within high-dimensional neural responses.

Our general theory allows population activity $\vr$ to encode the underlying message-passing model in a potentially complicated and nonlinear fashion, described as $\vr_t = \mathcal{R}\lpr \vx_t, \veta_t \rpr$, where $\mathcal{R}$ is the neural encoding function, $\vx_t$ is the collective state of all nodes at time $t$, and $\veta_t$ is noise or variability not related to the mental model. While we currently assume a simple linear encoding, the framework also generalizes to nonlinear encoding. Furthermore, we assume that the spatial pattern of neural activity instantaneously represents the information about $\vx_t$, although we could also consider alternative models in which probabilities are represented through temporal patterns \cite{savin2014spatio,deneve2008bayesian,hoyer2003interpreting} or spatiotemporal patterns \cite{lange2022interpolating}. In a spatial code, the population activity $\vr_t$ evolves such that the dynamics of the encoded node states $\vx_t$ conforms to the update equations (Eqs~\ref{eq:message_function}--\ref{eq:node_update_function}), thereby implicitly representing the dynamics of message-passing inference on the underlying graphical model \cite{raju2016inference,pitkow2017inference}.

Integrating these theoretical perspectives, we arrive at our core {\it neural message-passing hypothesis}: the brain approximates probabilistic inference on an internal model of the world using nonlinear message-passing between overlapping neuronal populations.

\subsection*{Framework for Inferring Inference}

To reveal the brain's inferential computations under this hypothesis, we develop an analysis framework --- {\it Inferring Inference} --- that can identify the latent graphical model and message-passing dynamics from stochastic neural data and sensory inputs. If our hypothesized structure is correct, a model fit by this framework will recover the nonlinear message-passing algorithm well enough to accurately predict neural population responses to novel inputs.

Given sensory observations $\vo_t$ and recorded neural population activity $\vr_t$ from a perceptual inference task, our aim is to simultaneously find: $\lpr i \rpr$ the neural encoding of relevant latent variables, $\lpr ii \rpr$ the interactions between these variables that define the probabilistic model, $\lpr iii \rpr$ the canonical message functions that specify the implicit algorithm, and $\lpr iv \rpr$ the mapping from sensory observations to network inputs. We achieve this by constructing a Hidden Markov Model (HMM) in which each of these elements affects the likelihood of the measured neural activity; inferring them then reduces to a maximum likelihood estimation problem.

To construct this HMM without ambiguity, we explicitly distinguish between two nested types of dynamic latent variables:
\begin{itemize}
\item {\bf Latent to the brain (world state):} As depicted in the mechanistic model (Fig~\ref{fig:Schematic}A), an evolving true world state $\vw_t$ generates sensory observations $\vo_t$. According to the algorithmic model, the brain uses these observations to infer the state of its own {\em assumed} task-relevant variables $\vs$, which according to the brain are related according to an assumed PGM (Fig~\ref{fig:Schematic}B, C). Specifically, we assume the brain infers approximate marginal posteriors $q_i$ for each node $i$ in this graph.
\item {\bf Latent to the scientist (algorithmic state):} Unknown to us are the brain's dynamic node states $\vx_{it}$ that describe the brain's marginal posteriors. We model the latent dynamics of $\vx_t$ using a message-passing algorithm (Eqs~\ref{eq:message_function}--\ref{eq:node_update_function}) and use them to account for experimentally measured neural activity time series $\vr_t$.
\end{itemize}

We collect all the components governing the dynamics of these states into a unified parameter set $\vtheta$. This set includes: the neural encoding function $\mathcal{R}$ that maps node states $\vx_{t}$ to population activity $\vr_t$, the input mapping function $\mathcal{V}$ from sensory observations $\vo_t$ to node inputs $\vv_{it}$, the structural couplings $\vJ_{ij}$, and the canonical message-passing operations $\mathcal{M}$, $\mathcal{A}$, and $\mathcal{U}$. By leveraging the Markovian nature of the message-passing dynamics, we can tractably estimate the parameters $\vtheta$ that best explain the measured neural responses via maximum likelihood.

In the following sections, we apply this framework to simulated neural data whose latent dynamics accord with the neural message passing hypothesis, and evaluate the ability of our framework to recover the ground truth as a function of various experimental properties. 

\begin{table}[ht!]
    \renewcommand{\arraystretch}{1.2}
    \centering
    \rowcolors{1}{white}{gray!12}
    \begin{tabular}{rrl}
        \hline
        &{\it Symbol} & {\it Meaning} \\
        \hline
        World model &$\vw$ & true world state \\
        &$\vs$ & brain's assumed latent world state \\
        &$\vo$ & sensory observation \\
        &$q(\cdot)$ & approximate posterior over assumed world state \\
        &$\vx$ & parameter of posterior \\
        &$t$ & time \\
        
        \hline
        Neural &$\mathcal{G}$ & graph of interactions between assumed latents \\
        message-passing &$\vx_{it}$ & vector-valued posterior parameter of node $i$ at time $t$ \\
        &$\vJ_{ij}$ & vector-valued model parameter of edge $(i,j)$ \\
        &$\vv_{it}$ & local input at node $i$ at time $t$ \\
        &$m_{j\to i, t}$ & message from node $j$ to $i$ at time $t$ \\
        &$\mathcal{M}$ & message function \\
        &$\mathcal{A}$ & message aggregation function \\
        &$\mathcal{U}$ & node update function  \\
        &$\mathcal{R}$ & neural encoding function  \\
        &$\mathcal{V}$ & input mapping function  \\
        
        \hline
        TAP model &$s_i$ & latent variable corresponding to node $i$\\
        &$J_{ij}$ & direct statistical interaction between nodes $i$ and $j$\\
        &$\vo_t$ & external input at time $t$\\
        &$V$ & input mapping matrix (linear form of $\mathcal{V}$) \\
        &$x_{it}$ & approximate marginal posterior of $s_i$ at time $t$\\
        &$\vx_t$ & vector of marginal probabilities at time $t$\\
        &$\lambda$ & update rate of the TAP dynamics \\
        &$\sigma(\cdot)$ & sigmoid nonlinearity \\
        &$\vxi_t$, $\Sigma_{\xi}$ & process noise and its covariance \\
        &$R$ & neural encoding matrix (linear form of $\mathcal{R}$) \\
        &$\vb$ & offset of the neural encoding \\
        &$\veta_t$, $\Sigma_{\eta}$ & measurement noise and its covariance \\
        &$\vr_t$ & population neural activity at time $t$\\

        \hline
        Model fitting &$p(\cdot)$ & probability of data under message-passing model \\
        &$G$ & canonical message-passing parameters \\
        &$G_{abc}$ & coefficient of the polynomial term $J_{ij}^a x_{it}^b x_{jt}^c$ \\
        &$K$ & number of particles in the particle filter \\
        &$\vtheta$ & vector of all parameters to be estimated $\lpr R, \vb, V, J, G \rpr$ \\
        \hline
    \end{tabular}
    \vspace{2mm}
    \caption{{\bf Glossary of notation.}}
    \label{tab:notation}
\end{table} 

\subsection*{A concrete model brain implicitly performing approximate inference}
To demonstrate our framework, we analyze artificial neural recordings generated by a simple model brain. We construct a brain that implicitly implements an advanced mean-field inference on a binary world state, following the Thouless-Anderson-Palmer (TAP) equation \cite{thouless1977} derived originally to describe disordered physical systems and subsequently used in a variety of applications \cite{opper2001tractable,murayama2004thouless,shamir2000thouless}. This example is a non-trivial instance of a hidden message-passing algorithm for approximate inference.

At a {\it normative} level, this inference model estimates marginal probabilities of $N_s$ binary latent variables, $\vs \in \lcr -1,1\rcr^{N_s}$, from the joint distribution $q\lpr \vs | \vo \rpr \propto \exp \lpr \vs^{\top}\!J\vs + \vs^{\top}V\vo\rpr$. Here $\vo$ are the sensory inputs, $V$ is a linear mapping from these inputs to the latent space, and $J$ is a coupling matrix that defines the graphical model, with $J_{ij}=0$ for states $s_i$ and $s_j$ that do not interact directly. 

At the {\it algorithmic} level, the latent state $\vx$ represents the approximate marginal probability. 
For the TAP equation, each assumed world state $s_i$ is just one binary variable, so we can summarize each marginal distribution compactly by just a single number, $x_i = q_i(s_i=+1|\vo)$, which specifies the approximate probability that the latent state is $+1$. The TAP model has nontrivial canonical nonlinear dynamics yet a relatively simple low-order polynomial form. These dynamics have a message-passing structure, where the node state $\vx_{it}$ is updated based on messages received from its neighbors and the sensory input $\vo_t$,
\begin{equation}
x_{i\,t+1} = (1-\lambda)x_{it} + \lambda\,\sigma\biggl( \sum_{j=1}^{N_s} \mathcal{M}\lpr x_{it}, x_{jt}, J_{ij} \rpr + \lpr V \vo_t \rpr_{i} \biggr) + \xi_{it}\hspace{3mm} i=1,...,N_s 
\label{eq:TAP_equation}
\end{equation}
Here, $\lambda \in (0,1]$ is a relaxation parameter setting the timescale for the dynamics, $\sigmoid(z)=1/(1+e^{-z})$ is a sigmoid function that affects the update operation (Eq~\ref{eq:node_update_function}), the aggregation function is a sum (Eq~\ref{eq:aggregation_function}), and $\mathcal{M}$ is a canonical message function (Eq~\ref{eq:message_function}) shared across the graphical model, specifically,
\begin{equation}
\mathcal{M}\lpr x_{i}, x_{j}, J_{ij} \rpr = 2J_{ij}x_{j} + 4J_{ij}^2\lpr 1 - 2x_{i} \rpr x_{j} \lpr 1 - x_{j} \rpr
\label{eq:TAP_message_function}
\end{equation}
Unlike the usual TAP equation, we also add a small amount of Gaussian process noise $\vxi_t$ to emulate the stochasticity found in brains' real dynamics and allow for some model mismatch.

\begin{figure}[ht]
	\centering
	\includegraphics[width=1\textwidth]{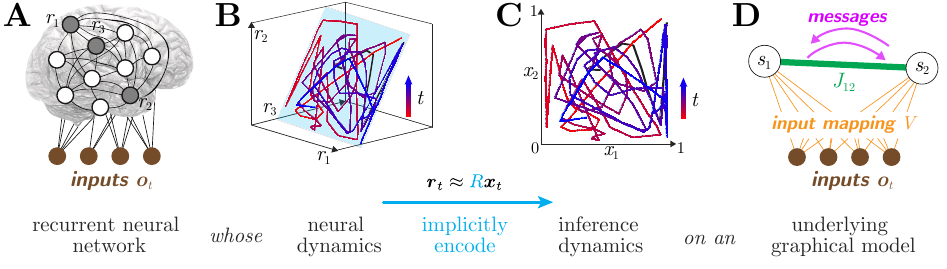}
	\caption{{\bf A model brain}, implemented as a trained RNN ({\bf A}), has neuronal dynamics ({\bf B}) that are an approximate linear embedding of the TAP inference dynamics ({\bf C}) on a binary probabilistic graphical model ({\bf D}). In this illustration, the joint activity of the three gray neurons traces out a trajectory (time indicated as red to blue) in a 2D subspace corresponding to the inference dynamics of two interacting variables.
	}
	\label{fig:TAPBrain}
\end{figure}

At the {\it mechanistic} level, the model brain that enacts this implicit inference algorithm is a two-layer recurrent neural network (RNN) with ReLU activations. 
We train the RNN such that the target low-dimensional latent dynamics are approximately linearly embedded by a matrix $R$ into the neural activity: $\vr_t \approx R\vx_t+\vb+\veta_t$ where $\veta_t$ is additive white Gaussian encoding noise. Thus, by construction, this simulated brain implicitly implements inference by message-passing on the underlying graphical model (Fig~\ref{fig:TAPBrain}).

\subsection*{Inferring inference in an example TAP brain}

Consider an example TAP brain with $N_r =500$ neurons that receives inputs of dimension $N_o = 10$ and encodes $N_s = 10$ latent variables.
The coupling matrix $J$ is a randomly generated sparse symmetric matrix, shown as a graphical model in Fig~\ref{fig:Example-simulation}D.
The neural encoding matrix $R$ and the input mapping matrix $V$ are also randomly generated, and both distribute their respective signals densely across all neurons. Random input signals $\vo_t$ 
evoke neural responses $\vr_t$.
Here we allow the measurement of all 500 neurons in the model brain, but we expect that if there were more neurons then they would be largely redundant with the ones we do measure, reflecting the same latent dynamics but with a higher signal to noise ratio \cite{gao2017theory}.

\begin{figure}[htbp!]
	\centering
	\includegraphics[width=\textwidth]{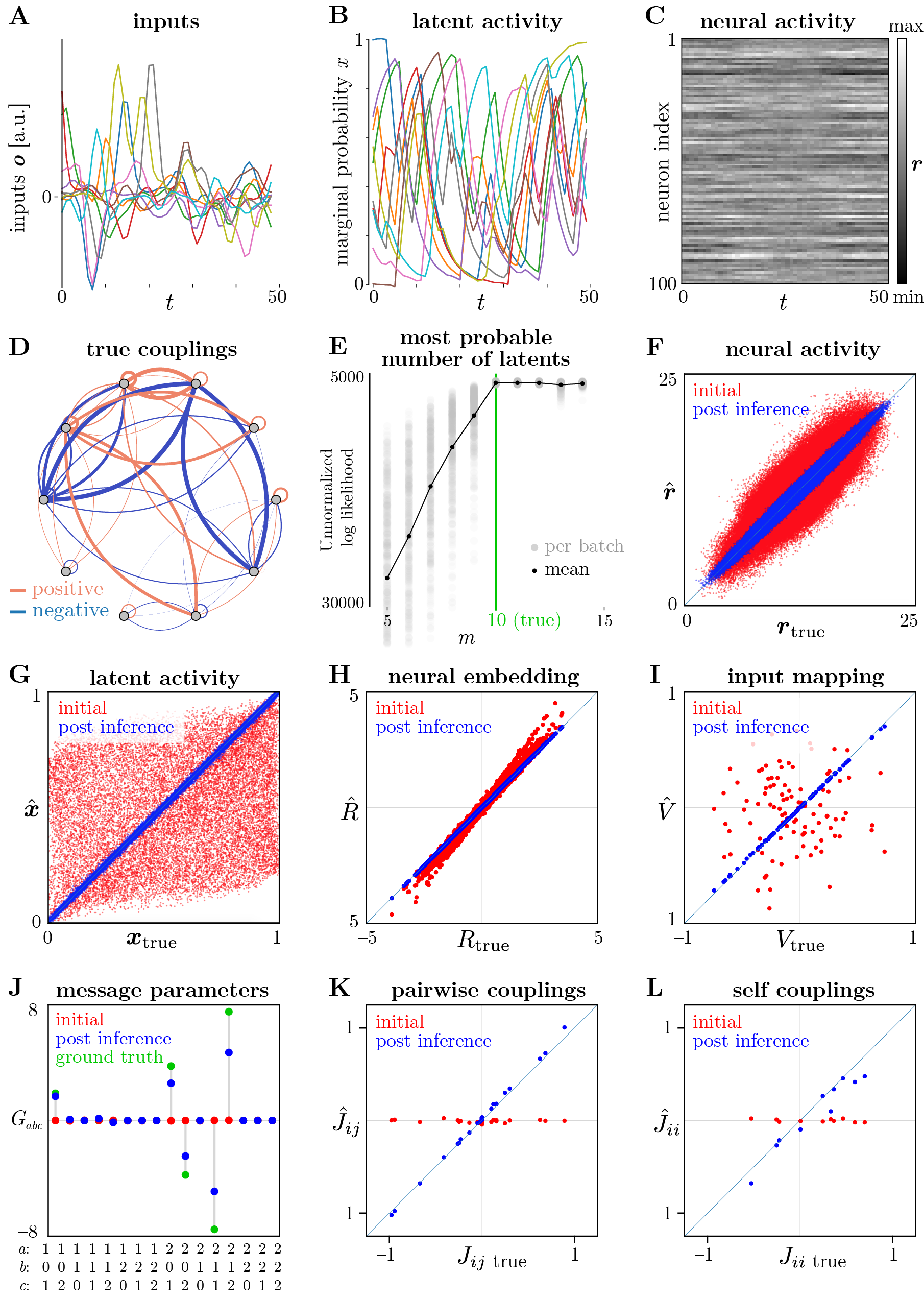}
	\caption{(Caption on next page.)}
	\label{fig:Example-simulation}
\end{figure}
\begin{figure}[t]
  \contcaption{
  {\bf Inferring inference in an example TAP brain}. The example brain has $N_r=500$ neurons that encode the dynamics of $N_s=10$ latent variables. We use Particle EM to fit the neural measurements and successfully recover the latent dynamics, followed by a backward greedy optimization to refine the estimates of the message parameters $\hat{G}$ and the coupling matrix $\hat{J}$. {\bf A}: Dynamic inputs $\vo$ to the model, used by the simulated brain to infer states of the external world. {\bf B}: The internal model's latent states $\vx$ are the approximate marginal probabilities $q$ of each world state according to the internal model. {\bf C}: These latent states are encoded in the distributed activity $\vr$ of neurons (100 of the 500 neurons shown). {\bf D}: The ground truth graphical model, where the edges correspond to couplings. Positive and negative couplings are colored orange and blue, respectively, and the edge thickness represents the relative magnitude. {\bf E}: Observed data log-likelihood on held-out test batches after Particle EM (gray: individual batches; black: mean), vs. the assumed number of latent variables. The likelihood is highest for the true number of latents, ($\hat{N}_s = 10$, green line). {\bf F}: Scatter plot of the fit to neural activity vs. the ground truth. {\bf G}: Scatter plot of the inferred latent activity vs. the ground truth. In panels {\bf F} and {\bf G}, each data point corresponds to one time sample of one neuron/latent variable. The red and blue data correspond to the initial and post-inference values, respectively. {\bf H} and {\bf I}: Scatter plot of the estimates vs. ground truth values of the elements of the neural embedding and input mapping matrices, respectively. {\bf J}: The true (green), initial (red), and final estimates (blue) values of the canonical message parameters $G_{abc}$.  {\bf K} and {\bf L}: Scatter plot of the estimates vs. ground truth values of the pairwise coupling and self-coupling (diagonal) terms of $J$.}
\end{figure}

Given this model brain, we would now like to infer its parameters using only the measured neural activity and sensory inputs.

To recover the latent dynamics given sensory inputs $\vo_t$ and measurements $\vr_t$ from a simulated brain, we instantiate the general framework with the following simplifying assumptions: 
$\lpr i \rpr$ the neural activity is a linear embedding of the latent dynamics, $\lpr ii \rpr$ the aggregation function in the message-passing dynamics is a summation, $\lpr iii \rpr$  the mapping from inputs to the latent space is linear, and $\lpr iv \rpr$ the latent dynamics have noise $\vxi_t$ and the encoding has noise $\veta_t$, both Gaussian-distributed with known or estimable covariances. Our goal is thereby reduced to simultaneously estimating
the linear embedding matrix $R$, linear mapping from inputs to the latent space $V$, the coupling matrix $J$ that defines the graphical model, and the parameters of the canonical message and update functions. Since the parameters to be estimated also depend on the unobserved latent variables, we use the Expectation-Maximization (EM) algorithm \cite{dempster1977EM} for maximum likelihood estimation. 

The E-step of the EM algorithm requires computing the posterior distribution of the latent variables, a challenging inference problem in models with nonlinear latent dynamics. Here we used a particle filter \cite{cappe2007overview,kantas2015particle}, also known as Sequential Monte Carlo (SMC), to flexibly approximate the posterior over latent states as a point cloud of sampled state trajectories $\vx_t$. 
This iterative combination of particle filters with EM for estimating unknown parameters in latent variable models is known as Particle EM \cite{cappe2007overview} (details in Methods). We apply this approach to simulated neural measurements to obtain the maximum likelihood estimate of the underlying parameters.

For our initial analysis, we make the simplifying assumption that the form of the update function is known. We also assume (correctly, for this example) that the unknown canonical message function $\mathcal{M}$ can be expressed in a low-order polynomial basis as
\begin{equation}
\mathcal{M} \lpr x_i, x_j, J_{ij} \rpr = \sum_{a,b,c}G_{abc}J_{ij}^a x_{i}^b x_{j}^c
\label{eq:general_message_function}
\end{equation}
where $a,b,c$ are integer powers of monomial terms. The corresponding coefficients $G_{abc}$, which we call the canonical message parameters, are global (shared across the entire graphical model) and define the specific nonlinearity of the message function. We relax these two assumptions in a subsequent section.

We obtain an initial estimate of the neural embedding $\hat{R}$ using Independent Component Analysis (ICA) \cite{hyvarinen2000independent} on measured neural activity, since ICA is particularly well-suited to discovering non-Gaussian dimensions. 

Right away we are faced with the challenge that the true number of latent variables encoded by the brain is unknown a priori. We address this as a model selection problem: we apply the Particle EM algorithm multiple times, iterating over a range of assumed latent variable counts, $m$. For each assumed value of $m$, we estimate the optimal parameters and then select the value of $m$ that yields the highest likelihood, averaged across a held-out set of test data.
This procedure reliably identifies the correct number of latent variables in our application (Fig~\ref{fig:Example-simulation}E). 

Once the number of latent variables has been identified, we can examine the inference solution with the highest likelihood.  Fig~\ref{fig:Example-simulation}F and G compare the model fit and the ground truth, for both neural measurements and the latent dynamics using previously unseen test inputs. The red and blue data points correspond to our initial and final estimates of parameters. 
After Particle EM converges, these estimated parameters reliably recover the ground truth latent dynamics. Fig~\ref{fig:Example-simulation}H,I shows that our method recovers accurate estimates of both the neural encoding matrix $\hat{R}$ and the input mapping $\hat{V}$ matrix.

Note, however, that the inferred coupling matrix $\hat{J}$ and message passing parameters $\hat{G}$ differ from their true value, even though the model provides an excellent fit to the observable data. This is a consequence of the degeneracies in our parameterization \cite{genkin2020moving}. To break these degeneracies, we impose a sparsity regularizer, progressively pruning $G$ to find a sparse subset of message-passing parameters that best explain the latent dynamics (see \nameref{subsection:methods_greedy_optimization} in Methods). Fig~\ref{fig:Example-simulation}J--L shows that after this refinement step, the inferred message-passing parameters and the couplings closely match the ground truth. The small discrepancy that remains in \ref{fig:Example-simulation}J reflects a degeneracy that our regularization cannot resolve. 
Notice that this discrepancy does not appreciably affect the inference, since the latent variables are nonetheless inferred correctly (\ref{fig:Example-simulation}G).
With suitable regularization, we are thus able to infer the implicit inference computations of our model brain.

\subsection*{Better experimental design for better inferences}
\label{sec:experimental_design}

An important factor for the success of our analysis framework is experimental design, {\it i.e.} choosing stimuli or tasks that can reveal the relevant encoding and nonlinear dynamics. Strong stimuli may drive latent variables to extremes of their range, which makes it easier to identify the dimensions that encode these variables. However, these same inputs may be so strong that they overwhelm the effects of recurrent interactions, which makes it difficult to learn these interactions. Conversely, weak stimuli may bring a network to an approximately linear regime where the interactions are distinguishable from input but not from each other: weak stimuli may not expose the interesting nonlinear interactions that characterize different computational algorithms. We found that it is best to present a distribution of stimuli with a wide range of intensities. This allows us to identify both the embedding and the dynamics.

\begin{figure}[ht!]
\centering
\includegraphics[width=\textwidth]{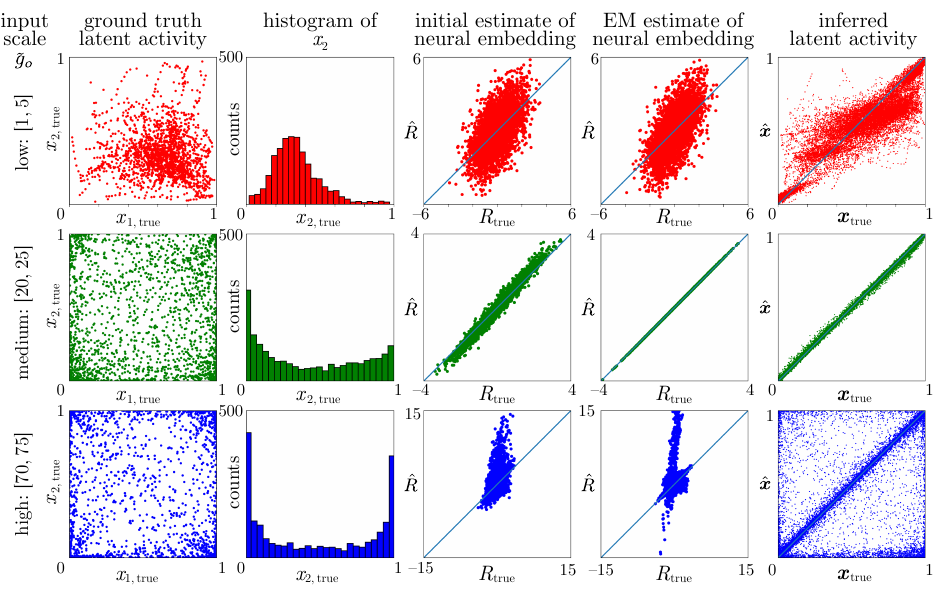}
\caption{{\bf Importance of experiment design for inferring latent structure.} Each row corresponds to a different range of amplitude scaling factor of the inputs $\tilde{g}_o$ relative to the coupling strengths. The top, middle, and bottom rows correspond to low, medium, and high input gain ranges, respectively. The first column from the left shows a scatter plot of two components of the ground truth latent activity $\lpr x_1, x_2 \rpr$ of the example TAP brain in Fig~\ref{fig:Example-simulation}. The second column shows the histogram of one component of the ground truth latent activity $\lpr x_2 \rpr$. The third column from the left shows scatter plots of the initial ICA estimate of the neural embedding matrix vs. the ground truth. Similarly, the fourth column corresponds to the EM estimate of the neural embedding. The rightmost column shows a scatter plot of the inferred vs. the ground truth latent activity. When low-gain inputs (top row/red) are used, the latent dynamics are in a localized regime and the resulting ICA estimate $\hat{R}$ is poor. This leads to degenerate solutions with Particle EM. On the other hand, when the input gain is too high (bottom row/blue), the latent activity is more biased towards maximal and minimal values. The resulting ICA estimate of the neural embedding $\hat{R}$ has large deviations from the ground truth along multiple dimensions, again leading to suboptimal inference of the latent dynamics. However, when the input gain is the range $[20, 25]$ (middle row/green), the latent dynamics exhibit a fairly uniform distribution of the latent states. In this regime, ICA is able to obtain a good initial estimate $\hat{R}$, with which Particle EM correctly estimates the neural encoding dimensions and the latent activity. 
}
\label{fig:ExperimentDesign}
\end{figure}

We illustrate this in the context of our model brain as follows. The quality of our initial parameter estimate depends on the strength of the input signal relative to the coupling strengths. To control the input strength, our experimental stimuli use an amplitude scaling factor $\tilde{g}_o$ to quantify the relative input gain (see \nameref{subsection:methods_input_signal_design} in Methods for details). 

Fig~\ref{fig:ExperimentDesign} shows how the ground truth latent activity, initial estimate of the neural embedding, and the Particle EM estimates each vary as a function of this gain for the example TAP brain in Fig~\ref{fig:Example-simulation}.   
In the case of low gain inputs (Fig~\ref{fig:ExperimentDesign}, top), the weak constraints result in ICA estimates that have large deviations from the ground truth. Extremely high gain inputs are not ideal either (Fig~\ref{fig:ExperimentDesign}, bottom); the latent activity is biased towards maximal or minimal values and we observe that ICA fails to recover a good estimate of the encoding matrix $\hat{R}$ along multiple dimensions. When $\tilde{g}_o$ is in the range $[20, 25]$, we observe that the latent dynamics explore their full range, resulting in a fairly uniform distribution of the latent states over time. Under this condition, ICA is able to obtain a reliable neural embedding $\hat{R}$ (middle row, third column). With this initial estimate, the Particle EM algorithm is biased towards the correct subspace for the latent dynamics (Fig~\ref{fig:ExperimentDesign}: middle row, fourth and fifth columns from the left in Fig~\ref{fig:ExperimentDesign}). For the inference results with the example TAP brain in Fig~\ref{fig:Example-simulation}, we used a broad range of input strengths $\tilde{g}_o \sim \mathcal{U}[5, 25]$ to enable the model brain to exhibit a wide repertoire of its dynamics, and this enabled us to recover the true internal model.

\subsection*{Importance of using the correct inductive bias}
\label{sec:correct_inductive_bias}
The Inferring Inference framework is predicated on the core hypothesis that the underlying inference process operates via a nonlinear message-passing algorithm on a structured graph. In addition, in the results presented in Fig~\ref{fig:Example-simulation}, we used a polynomial message function (Eq~\ref{eq:general_message_function}) that was precisely matched to the ground-truth dynamics, and we assumed that the form of the aggregation and update functions were known {\it a priori}. To assess how necessary these assumptions are, we systematically relaxed them, using generic Multi-Layer Perceptrons (MLPs) to model: $(i)$ only the message function; $(ii)$ both the message and update functions; and $(iii)$ the complete latent state transition using a single, monolithic MLP. This final variation effectively abandons the core assumption of a sparse graph structure altogether.

\begin{figure}[htb!]
    \centering
    \includegraphics[width=\linewidth]{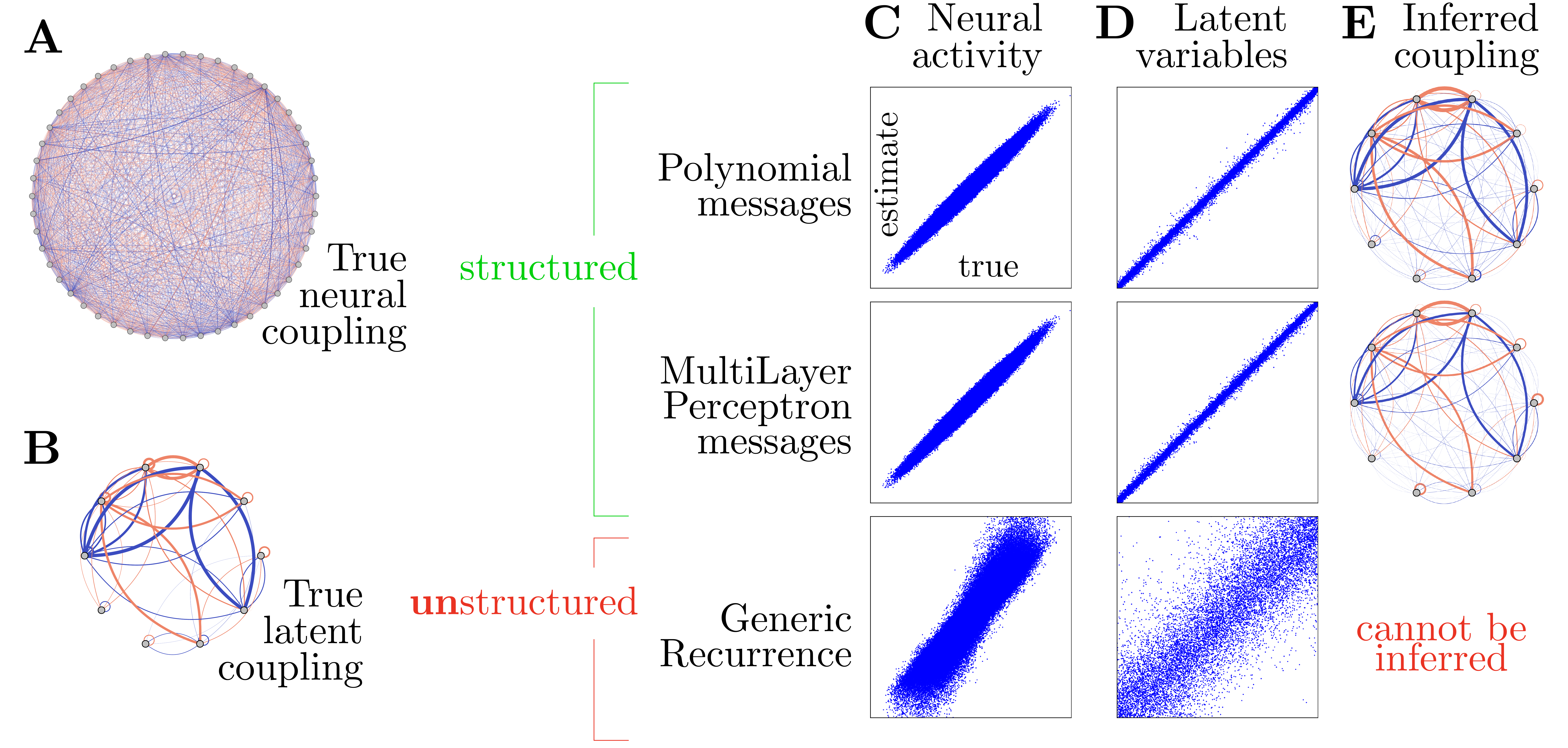}
    \caption{The importance of using the correct inductive bias. The TAP brain, from Fig~\ref{fig:Example-simulation} (with dense, recurrent connectivity shown in {\bf A}, visualized using a subset of 50 out of 500 total neurons) was trained to possess implicit low-dimensional nonlinear graph-structured interactions (with interactions shown in {\bf B}). Panels {\bf C}, {\bf D} and {\bf E} show results of our inference methods under different assumptions about the nonlinear dynamics: a polynomial message function (top, `Polynomial messages': a representation that includes the true latent dynamics), an MLP message function operating over the latent graph (middle, `MultiLayer Perceptron messages': closely related to Graph Neural Networks \cite{scarselli2009graph}), and MLP latent dynamics, a generic all-to-all nonlinear dynamical system without graph structure (bottom, `Generic Recurrence'). Fig~\ref{fig:messagePassingGeneralization-full} shows all four model variants, including MLP message and update functions. Given enough data, {\it all} of these methods can accurately reproduce the targeted neural activity and the underlying latents (not shown). However, with {\it limited} data (and neuroscience data is inevitably severely limited), the two structured dynamics models substantially outperform the unstructured MLP latent dynamics model at both reproducing neural activity ({\bf C}) and latent variables ({\bf D}) --- even when these inputs and trajectories have never been seen before (as shown here). Here we used only 100 sequences with 15 time steps for this inference. Moreover, the structured models provide interpretability through their connectivity ({\bf E}), whereas a generic dynamical system does not have explicit structure.}
    \label{fig:messagePassingGeneralization}
\end{figure}

\begin{figure}[htb!]
    \centering
    \includegraphics[width=\linewidth]{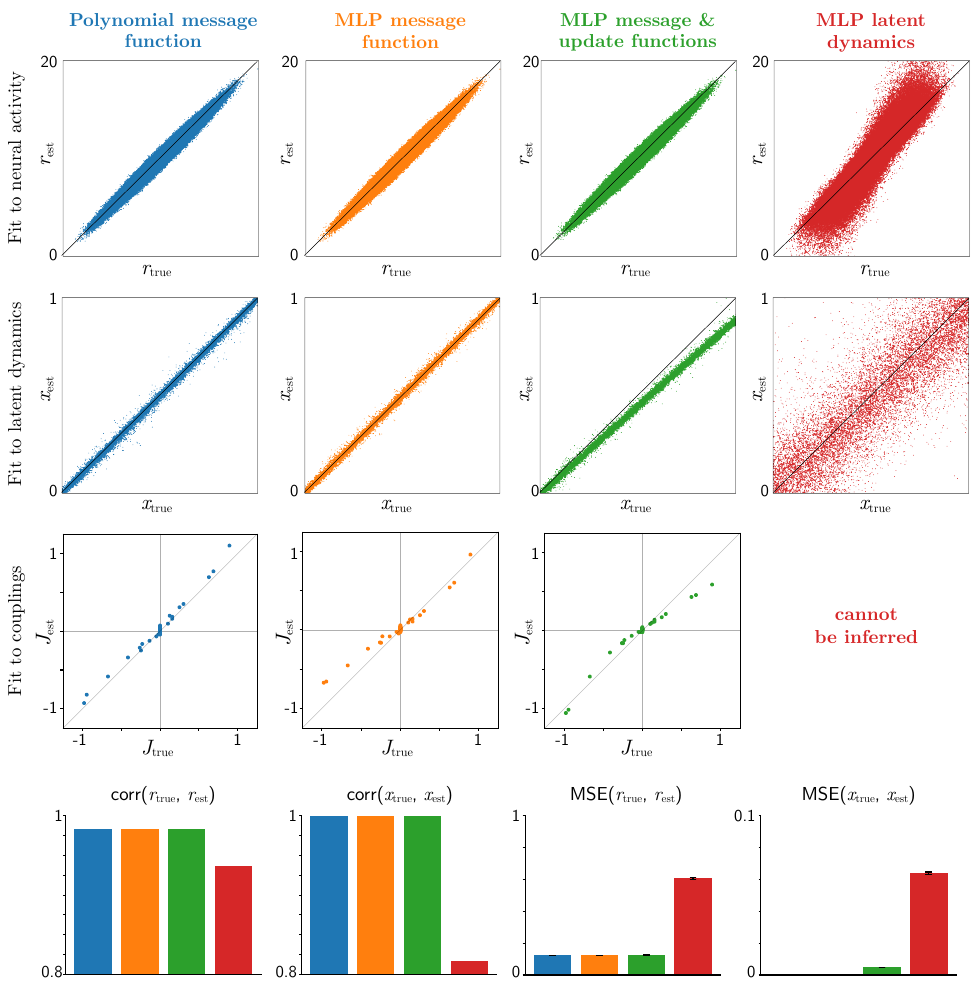}
    \caption{{\bf Inference performance of all four latent dynamics models in the data-limited regime.} Columns: polynomial message function, MLP message function, MLP message and update functions, and MLP latent dynamics (see Materials and methods). Rows 1--3: estimated versus true neural activity, latent variables, and couplings $J_{ij}$; couplings are not defined for the MLP latent dynamics model. Bottom row: correlation and mean squared error (MSE) between estimated and true neural activity ($\vr$) and latent variables ($\vx$).}
    \label{fig:messagePassingGeneralization-full}
\end{figure}

When provided with an abundance of data, we observed that all these latent dynamics models -- from the structurally correct model to the fully generic MLP dynamics -- could be used with Particle EM to achieve a good fit to the observed neural activity. However, we found this is no longer true in the limited data regime, as is invariably the case with neuroscience data. Fig~\ref{fig:messagePassingGeneralization} clearly demonstrates that employing the correct inductive bias, which in this case is message passing dynamics on a graph-structured model, fundamentally outperforms generic dynamics (see Fig~\ref{fig:messagePassingGeneralization-full} for all four model variants). This result highlights that while generic function approximators can eventually converge to the true dynamics with sufficient data, incorporating structural knowledge about the underlying computational mechanism allows for a highly constrained hypothesis space.

\section*{Discussion}

We set out to meet the ambitious goal of recovering a canonical algorithm only from neural data and sensory inputs. At first glance this seems impossible or at least ill-posed, but we showed that we could successfully recover structured latent message-passing dynamics in a distributed, multiplexed code, even when we don't know the latent variables, structure, or algorithm in advance. We fit this model using time series of simulated neural activity and a set of external inputs, and a few assumptions about the underlying computation. Although these assumptions were significant, as described below, this work provides an important proof of concept for Inferring Inference, and subsequent studies can relax these assumptions for greater generality.

\subsection*{Related work}

Past efforts to interpret neural activity have generally taken two complementary approaches: relating neural responses to task variables, or summarizing neural activity by low-dimensional latent variables. In the first case, task-relevant variables typically include sensory stimuli \cite{hubel1962receptive} and motor actions \cite{churchland2012neural}, but occasionally also include human-named latent variables like value or confidence \cite{glascher2010states,daw2005uncertainty,yu2005uncertainty,starkweather2017dopamine,ma2006bayesian,walker2020neural,henaff2020representation,wu2020rational}. In contrast, dimensionality-reduction methods are often claimed to be interpretable simply due to their lower dimensionality compared to the full neural dimensionality \cite{williamson2019bridging,cunningham2014dimensionality,glaser2020recurrent,duncker2019learning}. Our work combines these two approaches by identifying representations of latent variables while also attributing meaning from their structured statistical interactions.

Given our foundational model assumptions, this combination led directly to a computationally constrained statistical neural data analysis \cite{turner2013bayesian,durstewitz2016computational,linderman2017using,mlynarski2021statistical}.  
Other analysis methods share this aim of discovering latent computational functions from data, as opposed to a phenomenological description of the data. One prominent group of such studies includes inverse reinforcement learning \cite{ng2000algorithms,babes2011apprenticeship}, inverse optimal control \cite{dvijotham2010inverse,schmitt2017see,straub2022putting}, and inverse rational control \cite{wu2020rational,kwon2020inverse,jarrett2021inverse}. However, these approaches provide interpretability of behavioral data, not neural data as we use here. Some approaches attempt to jointly model behavior and neural data \cite{turner2013bayesian,turner2017approaches}, using the same latent variables for both. Perhaps the closest to our approach philosophically is \cite{chalk2021inferring}, which aims to discover optimality principles of a neural network by analyzing neural activity.

Our approach to discovering computations is premised upon structured dynamics. Here we aim to discover inferential dynamics, but other approaches aim to discover a wide variety of other types of parsimonious interactions.
Some of these use constrained nonlinear dynamics operating directly on observables, whereas others allow the dynamics to be hidden behind a subset of imperfect observations.

The past studies that concentrated on directly observable variables often have applications of inferring physical laws \cite{battaglia2018relational,schmidt2009distilling,baddoo2023physics}, biochemical reaction networks \cite{shindo2018inferring}, neural connectivity \cite{mi2022connectome} and neural properties \cite{allier_graph_2026}, or interactions between more general objects \cite{kipf2018neural}. These methods differ from ours because they identify structured interactions amongst predefined variables, whereas we also discover the latent variables themselves and their embedding in neural activity.

Other graph-based relational learning approaches aim to learn the latent variables, like our approach, but many localize the representation of each variable within a dedicated population \cite{sabour2017dynamic,scarselli2009graph,xi2017capsule,goyal2021recurrent}. This approach has been used for learnable graph-based inference in probabilistic graphical models \cite{yoon2018inference,fei2024higher,qu2019graph,li2022learning} and for structured world models \cite{kipf2020contrastive}.
A crucial difference between our approach and these past models is that we allow structure to be multiplexed across population activity \cite{raju2016inference,rigotti2013importance,mante2013context,maheswaranathan2019universality}.

Ample other work also aims to discover latent dynamics within neural activity or other observables. These use different assumptions about the linearity of the assumed dynamics and of the embedding. The most common approach is to model latent nonlinear dynamics as embedded linearly within observations. Our proposal falls into this category, although below we describe possible generalizations that relax this restriction. The most similar work to ours is perhaps \cite{langdon2025latent}, who discover low-dimensional circuit structure within neural manifolds. Another example of this approach is switching linear dynamical systems, which approximate nonlinear dynamics by a piecewise linear dynamics \cite{ghahramani2000variational,fox2008nonparametric,linderman2017bayesian,giahi2019dynamic}. Other methods express observations as linear combinations of smooth functions for the nonlinear dynamics, via such mechanisms as recurrent neural networks \cite{jaeger2004harnessing,pandarinath2018inferring} or Gaussian processes \cite{duncker2019learning,wang2005gaussian}.

An alternative approach discovers linear dynamics within nonlinear embeddings. The Koopman operator \cite{koopman1931hamiltonian} demonstrates that a sufficiently high-dimensional nonlinear embedding, such as a time-delay embedding \cite{takens1981detecting,brunton2017chaos,schmid2010dynamic}, can transform nonlinear dynamics into linear dynamics, as used in models of neural population activity \cite{gao2016linear} and in robotics and control \cite{bruder2019nonlinear,abraham2019active}. Fewer studies examine nonlinear dynamics in nonlinear embeddings, because these models are generally underconstrained. However, suitable regularization can provide enough structure to fruitfully fit nonlinear dynamics within curved data manifolds. For example, Bakarji et al.~\cite{bakarji2023discovering} use sparse identification of nonlinear dynamics (SINDy)~\cite{brunton2016discovering} within a learned nonlinear embedding. While these latent variable dynamics can involve components with specific subsets of latent variables, previous approaches do not have inductive biases that favor graph-structured interactions with canonical functions. This additional structure in our approach may provide greater interpretability than these other more generic latent variable methods, and might help provide a better match to internal modeling of causality in the world that we and others hypothesize is a core element of cognition \cite{pitkow2017inference,battaglia2018relational,goyal2021recurrent,pearl2018book}.

Future work may find it fruitful to merge behavioral and latent algorithmic models using both behavioral and neural data. Such an approach could provide constraints on structured internal models that define distributed computations in behavioral tasks \cite{turner2013bayesian,linderman2017using}, and could infer how computations are decoded to generate actions \cite{haefner2013inferring,pitkow2015can,cumming2016feedforward,yang2021revealing}.

\subsection*{Limitations and generalizations}


Although our method of inferring inference discovers a surprising amount of structure, it still makes significant assumptions that would be good to relax in future incarnations.

Our foundational assumption is that the brain computes via canonical message-passing on graphs. This imposes an algorithmic symmetry that distinguishes {\em generic} nonlinear dynamics from {\em structured} nonlinear dynamics.
In some sense this assumption is trivially true, if nodes are neurons and the graph is the anatomical connections between them \cite{allier_graph_2026}. Notably, we did {\em not} assume that the computational graph corresponded to an anatomical connectivity graph in any way, although we could use such data from large-scale functional connectomics \cite{bae_functional_2025} to constrain our models \cite{mi2022connectome}. However, just as in statistical physics, it may be that a simpler macroscopic picture emerges from the collective behavior of many microscopic elements \cite{saxena2019towards,hoel2013quantifying}.
Ultimately it is an empirical question whether our hypothesis of canonical, low-dimensional message-passing computation parsimoniously describes brain computation \cite{pitkow2017inference,AtomsofNeuralComputation}. To evaluate this hypothesis on real data, it will be helpful to compare the performance of our model to that of other models described above, and even compare to relaxed versions of our own method without canonical (shared) message functions.

We also made model-specific assumptions that simplified this difficult inference problem. These can be grouped into assumptions about the brain's internal model, the neural representation of that model, and the class of dynamics.

\subsubsection*{Assumptions about the brain's internal model}

Our strongest (and arguably worst) assumptions restricted the class of probabilistic graphical models being used for inference.
In particular, our method assumed --- correctly in this case --- that the synthetic brain assumed that the latent world states were binary, so its beliefs were marginal probabilities. Technically, any generic graphical models can be approximated by a binary one \cite{Jeb14}, but this may require unduly complex augmentation. It would often be more natural to consider richer classes of graphical models accounting for multidimensional and/or continuous variables \cite{li2022learning,mardia2008multivariate}. Additionally, here we only considered pairwise interactions, so our method would require some generalization to accommodate richer internal models that include multi-way interactions between variables \cite{fei2024higher,kschischang2001factor,zhang2020factor}. 
Next, among pairwise graphical models, we considered only undirected, static models. It should be fairly straightforward to generalize our approach to accommodate directed acyclic graphs that can capture causal structure. This could include models with Markovian latent dynamics, allowing inferences to actually use sequences of observation (thus far we did allow sequences of observations, but the modeled inferences were static, with no temporal predictions for future observations). 

\subsubsection*{Assumptions about the neural representation}

The assumptions of binary latent variables and linear embeddings made it easier to identify the neural representation --- particularly the manifold in which the dynamics operate --- because marginal probabilities lie within a hypercube. When good experimental design exposes the full dynamic range of the inferred marginals, the bounded edges of this hypercube structure create a platykurtic distribution within neural activity that can be discovered by linear Independent Components Analysis, even without accounting for the dynamics. If the brain linearly represents the same marginals in a different manner, for example as log-probabilities, then we could in principle identify either a nonlinear embedding of the marginals \cite{bakarji2023discovering,archer2015}, or a linear embedding of the log-marginals \cite{walker2020neural}. Our work can be generalized to accommodate such nonlinear embeddings, although this will introduce additional degeneracies between the inferred embedding and inferred dynamics. For example, a message function that combines independent evidence can be expressed as products of probabilities, or sums of log-probabilities.
Some of these degeneracies may be broken by assuming additional constraints on the nonlinear embeddings, while other degeneracies will remain \cite{hyvarinen1999nonlinear}.

\subsubsection*{Assumptions about dynamics}

Our results showed that graphical model structure is a key property that reduces representational degeneracies. Even for flexible parametric or sampling-based representations of joint distributions, sparse graph-structured interactions between variables can provide a core computational constraint.
Graph-structured interactions are a foundation of causal models as well, since graph assumptions are used to draw conclusions that go beyond the pure data \cite{pearl2009causality}.

While arbitrary graphs could in principle be paired with arbitrarily complex message functions like look-up tables to overfit observable data from a mismatched sparse graph, the erroneously inferred interactions will not generalize. Thus, to learn sparse graphs, we must constrain the message functions so they are not arbitrarily complex. 

An alternative approach to modeling dynamics is to use an embedding space that is sufficiently high dimensional that all nonlinear dynamics in the original space can be expressed as linear dynamics in the embedding space \cite{koopman1931hamiltonian,schmid2010dynamic,gao2016linear}. Although some of the graph structure could be preserved as block-structured linear dynamics, such a method must use a higher-dimensional state space that could make interpretation much more difficult.

Even though our model of the brain does include dynamics, those are only inference dynamics in a world that is assumed to be static. While we do allow the external evidence to change over time, the current underlying inference model treats these as noisy or mismatched observations rather than reflections of a changing world state. Subsequent versions of Inferring Inference should address this limitation by allowing an internal model based on a spatio{\it temporal} graph: a directed, time-translation-invariant Markov chain with internal spatial structure. This is a natural format for graph-structured causal variables describing a dynamic world.
This could arise from a hierarchical model of activity, and/or by continually learning interactions. Our framework could be augmented to account for these effects. 

Here we assumed the brain used essentially deterministic message-passing dynamics, although we allowed some small non-computational process noise. Alternative inference theories, notably sampling-based codes \cite{hoyer2003interpreting,lee2003hierarchical,ackley1985learning,berkes2011sampling,haefner2016perceptual}, use randomness for a computational function. Local sampling approaches, like Gibbs sampling, can be seen as stochastic message-passing, and our framework can be expanded to accommodate these dynamics as well. One interesting hybrid is to perform a temporal dimensionality reduction on the stochastic dynamics, smoothing time series while computing time-dependent nonlinear statistics at the fast timescale ({\it e.g.} slowly changing means and variances) that then flow through the graph via deterministic message-passing \cite{pitkow2017inference}.

\subsection*{Outlook}

To apply our framework to real data, we will need responses of many neurons responding to the same stimuli, and we will need a significant duration of data. To capture the message-passing dynamics, it may be important to record by fast techniques like electrophysiology, because slower techniques like calcium imaging could blur away the relevant computations. While the computations estimated on this slow timescale may still be nonlinear, they may hide faster message-passing dynamics that better explain these nonlinearities.

Depending on the properties of the code and our modeling assumptions, it may be important to have simultaneously recorded neurons. For example, in a sampling-based code, the joint uncertainty is reflected by response covariations \cite{hoyer2003interpreting,berkes2011sampling,haefner2016perceptual}.
Although these can be re-interpreted as nonlinear rate codes \cite{henaff2020representation,yang2021revealing} through expectation values \cite{vertes2018flexible}, estimating these nonlinear statistics still requires simultaneous measurements.

The notion of a canonical operation is, in a deep sense, core to {\em any} form of understanding: one rule that explains multiple computations; an {\em explanans} that is simpler than the {\em explanandum}. For other theories of brain computation, this core explanation might be a plasticity rule, a goal and optimization method \cite{richards2019deep,darwin2004origin}, and/or prewired structure \cite{zador2019critique,maass2002real,sinz2019engineering}.

How could the brain implement a canonical operation? Unlike machine learning algorithms, it is unable to copy synaptic weights to multiple locations during learning. We see three possibilities: hardwiring, convergent learning, and modularity. Hardwired local microcircuits \cite{mountcastle1997columnar} and control structures are conserved across animals, and these may arise from genetically encoded developmental programs selected by evolution \cite{zador2019critique,cisek2019resynthesizing}. Convergent learning could happen if a genetically encoded plasticity rule rediscovers a common algorithm multiple times across a graph because that is the optimal solution for many problems \cite{orhan2017efficient}. In between these extremes, it may be that system-wide architecture imposes bottlenecks that create a modular architecture \cite{dulberg2022modularity,logiaco2021thalamic,musslick2021rationalizing,zhang2024inductive}, so this module's computations can be `copied over time' by applying it sequentially to different inputs gated through the module. Our analysis approach could discover canonical computations, regardless of mechanism.

Although quite flexible, message-passing computation on graphs is not universal. It cannot, for example, distinguish between graphs and their covers (larger graphs that pass over the original nodes multiple times) \cite{chen2019equivalence,sato2020survey}. The class of message-passing algorithms could potentially predict specific inferential errors, like overcounting evidence --- predictions that we could directly test with suitably designed behavioral and neural experiments.

Modern neuroscience is acquiring massive amounts of data, such as the MICrONS project \cite{bae_functional_2025} and the Allen Brain Observatory \cite{de2020large}, providing new opportunities to understand brain function. To make sense of these massive datasets, we need principled mathematical frameworks \cite{pitkow2017inference,wu2020rational,chalk2021inferring,mlynarski2021statistical,friston2010free,ma2022principles}. The current approach of {\it Inferring Inference} may provide a novel way of identifying graph structure and canonical operations that the brain uses to model its environment and guide its actions.



\section*{Materials and methods}
\subsection*{Constructing a model brain that performs approximate inference}
\label{sec:construction}



\subsubsection*{Constructing a model brain}
\label{subsection:methods_constructing_TAP_brain}

The model brain is a two-layer recurrent neural network (RNN):
\begin{align}
\vr^\textrm{hid}_{t+1} &= \ReLU \lpr W^\textrm{hid}_\textrm{rec}\vr^\textrm{hid}_t + W^\textrm{hid}_\textrm{ff}\vo_t + \vb^\textrm{hid}\rpr \\
\vr_{t+1} &= \ReLU \lpr W_\textrm{rec}\vr_t + W_\textrm{ff}\vr^\textrm{hid}_t + \vb\rpr
\end{align}
where $\ReLU(x) = \max(0,x)$ is the Rectified Linear Unit activation function, $\vo_t$ is the input to the network, and $\vr^\textrm{hid}_t$ and $\vr_t$ are the activities of the hidden and output layers, respectively. $\{ W^\textrm{hid}_\textrm{rec}, W_\textrm{rec} \}$, $\{ W^\textrm{hid}_\textrm{ff}, W_\textrm{ff} \}$ and $\{ \vb^\textrm{hid}, \vb \}$ are the recurrent, feed-forward weights and biases of the RNN. This network is trained so the activity of the output layer is an approximate linear embedding of TAP dynamics on an underlying graphical model (eq. \ref{eq:TAP_equation}). 

The network has $N_h$ hidden layer neurons and $N_r$ output neurons. For our simulation example, we constructed a model brain with $N_r = 500$ output neurons, $N_h = 1000$ hidden layer neurons, and inputs of dimension $N_o = 10$. We chose this two-layer recurrent architecture because it was much easier to reproduce the desired latent dynamics than a single layer recurrent network. 
In all of our experiments with Inferring Inference for TAP model brains, we observe only the activity of the second layer neurons $\vr_t$. 

To construct this network, we must choose the coupling matrix that defines the assumed graphical model $J$, input mapping matrix $V$, and the neural embedding matrix $R$.
For the coupling matrix $J$, we generated a $N_s \times N_s$ sparse, symmetric adjacency matrix with sparsity of $0.5$. The non-zero elements of the coupling matrix were then sampled from $\mathcal{N}(0,1.5)$. For the input mapping matrix $V$, we used an orthogonal matrix obtained from the singular vectors in the singular value decomposition of a random $N_s \times N_o$ matrix with entries sampled independently from $\mathcal{N}(0,1)$. The elements of the $N_r \times N_s$ neural embedding matrix $R$ were also sampled independently from $\mathcal{N}(0,1.5)$.

For the chosen set of parameters and the time series of inputs $\vo_t$ described below, we generate the corresponding TAP dynamics $\vx_t$ using Eq~\ref{eq:TAP_equation}, providing a target, $R\vx_t + \vb$, for training the RNN. The constant bias vector $\vb$ is chosen to ensure that the target neural activity in the output layer is always positive. We train the network to minimize the squared error between the output neural activity and the target: $\sum_t \| \vr_t - R\vx_t - \vb\|_2^2$. For computing the MSE, we ignored the first $T_{\textrm{clip}} = 20$ time-steps of the RNN activity for each batch. We optimize the weights and biases of the RNN using the Adam optimizer\cite{kingma2015adam} with a learning rate of $10^{-5}$, mini-batch size of $16$, and $8\times 10^5$ training iterations.

\subsubsection*{Input signal design}
\label{subsection:methods_input_signal_design}
The choice of the input signals $\vo_t$ was crucial both for training the TAP model brain and indeed for applying the inferring inference analysis framework.

The inference algorithm enacted by the TAP equation assumes inputs are constant, and the dynamics in Eq~\ref{eq:TAP_equation} converge to a fixed point. To get more data about the underlying algorithm, we use dynamic input signals, contrary to the model brain's inference algorithm's assumptions. We chose these dynamic inputs to be filtered versions of random piecewise constant functions $\tilde{\vo}_t$ held constant for $T_{\textrm{const}}$ time-steps. At the start of each period, each input variable $\tilde{o}_i$ $(i = 1,...,N_o)$ is generated independently as $\tilde{o}_i = \gamma_i \nu_i$, where $\nu_i \sim \mathcal{N}(0,1)$. The amplitude scaling term is drawn from a Gamma distribution $\gamma_i \sim \Gamma\lpr\kappa, g_o\rpr$, where $\kappa=1$ is the shape parameter, $g_o = \tilde{g}_o/\sqrt{N_s}$ is the scale parameter, and $N_s$ is the number of latent variables in the TAP model brain. We then pass the raw input sequence $\tilde{\vo}_t$ through a forward-backward filter to obtain $\vo_t$. This is done to smoothen transitions between successive time periods. A normalized Hamming window of length $N_{\textrm{Ham}}$ is used as the impulse response of the filter. 
For each batch we sampled integer $T_{\textrm{const}}$ from the discrete uniform distribution $\mathcal{U}[2, 5]$. We set $N_{\textrm{Ham}} = 5$ as the Hamming window length for the temporal smoothing filter. 

When training the model brain, we used $\tilde{g}_o \sim \mathcal{U}[2, 50]$ for the continuous amplitude scaling distribution. We generated input signals for training the model brain with the following settings. We used $B_{\textrm{train}} = 25000$ training batches, each of length $T = 50$ time-steps. 
When inferring the algorithm of this model brain, we instead varied the amplitude as described in Fig~\ref{fig:ExperimentDesign}.

\subsection*{Inferring model parameters from simulated data}
\label{sec:methods_inferringinference}

\subsubsection*{Maximizing model likelihood}
\label{subsection:methods_particle_EM}

In this section we describe the Particle EM algorithm used to obtain the maximum likelihood estimate of the TAP brain parameters $\vtheta$, 
\begin{equation}
\hat{\vtheta} = \argmax \vtheta p \lpr \vr_{0:T}|\vo_{0:T};\vtheta\rpr = \argmax \vtheta \int p \lpr \vx_{0:T}, \vr_{0:T} |\vo_{0:T};\vtheta\rpr d\vx_{0:T}.
\end{equation}

The joint distribution of the latent and observed neural activity given the inputs is given by the Hidden Markov Model
\begin{equation}
p\lpr \vx_{0:T},\vr_{0:T}|\vo_{0:T};\vtheta\rpr = p(\vx_0)\prod_{t=0}^{T-1}p\lpr \vx_{t+1} | \vx_{t},\vo_{t};\vtheta\rpr \prod_{t=0}^{T} p\lpr \vr_t | \vx_t;\vtheta\rpr,
\end{equation}
where $\vtheta = \lpr R,\vb,V,J,G \rpr$ are the parameters to be estimated. The transition density and the emission (observation) density are specified as: 
\begin{subequations}
\begin{align}
p\lpr \vx_{t+1} | \vx_{t},\vo_{t}; \vtheta\rpr &= \mathcal{N}\lpr \boldsymbol{\mu} \lpr \vx_t,\vo_t ; \vtheta \rpr,\Sigma_{\xi}\rpr \\
p \lpr \vr_t|\vx_{t};\vtheta \rpr &= \mathcal{N}\lpr R\vx_t + \vb, \Sigma_{\eta}\rpr
\end{align}
\label{eq:methods-conditional-distributions}
\end{subequations}
where $\Sigma_{\xi}, \Sigma_{\eta}$ are the covariances of the process noise and measurement noise, respectively. 
The conditional mean of the transition probability is obtained by using the general polynomial form of the message function in Eq~\ref{eq:TAP_equation}, 
\begin{equation}
\mu_i \!\lpr \vx_t,\vo_t; \vtheta \rpr = (1-\lambda)x_{it} + \lambda\,\sigma\!\!\lpr \sum_{j,a,b,c}G_{abc}J_{ij}^a x_{it}^b x_{jt}^c +(V\vo_t)_i\rpr.
\label{eq:methods-conditional-mean}
\end{equation}

\noindent
A standard approach for computing maximum likelihood estimates of unknown parameters in models involving latent variables is the EM algorithm \cite{dempster1977EM}. However, the E-step requires us to compute the expected value of the complete data log likelihood with respect to the posterior distribution of the latent variables given the current estimate of the parameters $\vtheta_n$,
\begin{equation}
Q(\vtheta,\vtheta_n) \triangleq \mathbb{E}_{\vtheta_n} \lbr \log p \lpr \vx_{0:T}, \vr_{0:T}|\vo_{0:T}; \vtheta \rpr \rbr.
\label{eq:methods-E-step}
\end{equation}

\noindent
We use a particle filter to approximate the posterior distribution required in the E-step, and we use gradient ascent to perform the M-step (see \nameref{subsection:particle_EM} in Methods).

\subsubsection*{Greedy optimization of the message-passing parameters}
\label{subsection:methods_greedy_optimization}

\begin{figure}[ht]
	\centering
	\includegraphics[width=1\textwidth]{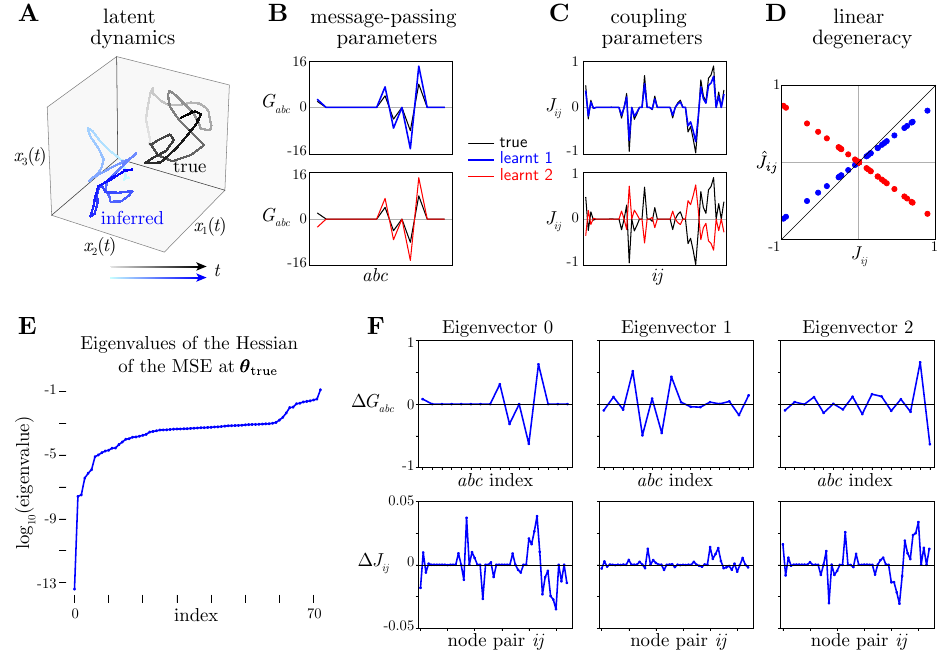}
	\caption[Illustration of system degeneracies]{{\bf Degeneracies in our parameterization.} {\bf A}: Latent states decoded by a particle filter (blue) versus true latent states (black). This visualization shows that the decoded trajectory is a linear transformation of the true trajectory. {\bf B,C}: If we fix the embedding to the correct value, and relearn the remaining parameters twice, then the two sets of parameters (red, blue) both provide reasonable matches to the message-passing parameters and couplings of the true model (black). Couplings $\hat{J}_{ij}$ (estimates) and $J_{ij}$ (true) are related by a linear transformation for each of the two solutions, shown in panel {\bf D}. $\hat{J}_{ij}$ in panel C and the corresponding message-passing parameters $\hat{G}$ in panel B result in an inference algorithm that is equivalent to the true model. {\bf E}. Log-eigenvalues of the curvature (Hessian) of the mean squared prediction error for the latent dynamics at the true parameter values. Small values indicate degeneracies. {\bf F}. The $G$ and $J$ components (top and bottom rows, respectively) of the eigenvectors corresponding to the three smallest eigenvalues. Parameter perturbations along eigenvector 0 correspond to the linear degeneracy between the interactions and the message passing parameters seen in B--D.
 }
	\label{fig:Degeneracies}
\end{figure}

We observe two broad classes of degeneracies in our optimization. The first class involves the neural embedding. We can recover latent dynamics that are just a linear transformation away from the true latent variable dynamics, $\hat{\vx}_t \approx A\vx_t$, in a way that is exactly compensated by a change in the neural embedding, $\hat{R} = RA$ (Fig~\ref{fig:Degeneracies}A).
The second class of degeneracies involves both the coupling parameters $J$ and the canonical message parameters $G$. Here, we recover the correct latent representations and dynamics, yet the estimates $\hat{J}$ and $\hat{G}$ differ from the ground truth. One simple version is that we can scale all the coupling terms $J_{ij}$ globally by any factor $\beta$, and then perfectly compensate by scaling each message-passing coefficient $G_{abc}$ by $\beta^{-a}$. This is equivalent to increasing the interaction energies and inference `temperature' at the same time, and leads to identical latent dynamics. Fig~\ref{fig:Degeneracies}B--D illustrates a typical degeneracy of this type.

To quantify the full range of degeneracies locally, we compute the curvature matrix (Hessian) of the mean squared prediction error at the ground-truth parameters, and compute the eigenvalues and eigenvectors of this Hessian (Fig~\ref{fig:Degeneracies}E,F). Eigenvectors with small eigenvalues indicate directions of low curvature, where prediction quality is similar as the model parameters change. Fig~\ref{fig:Degeneracies} shows several directions with small eigenvalues of the curvature matrix, revealing that there are several other interesting degeneracies between the interactions $J$ and message parameters $G$, all of which give rise to very similar latent dynamics.

Inferring the true latent representations and dynamics requires us to break these degeneracies in the parameterization. One potential approach is to use prior knowledge about the brain's internal model and introduce the appropriate inductive biases in our optimization framework. 
For instance, to break the degeneracy in the neural embedding, we might assume the latent states of the model brain are bounded from above and below, consistent with our assumed sigmoidal update function $\mathcal{U}$. 
This favors using Independent Component Analysis \cite{comon1994independent} to initialize the estimated neural embedding matrix $\hat{R}$, since it is effective at discovering embeddings of platykurtic distributions. 

To break the degeneracies in estimates of the coupling matrix $\hat{J}$ and the message-passing parameters $\hat{G}$ obtained from Particle EM, we find the smallest subset of $\hat{G}$ that best fits the inferred latent dynamics. This can be formulated as a minimization problem with $\ell_0$ regularization, adding an $\ell_0$-regularization term $\alpha||G||_0$ to the loss. General $\ell_0$ optimization is combinatorially hard, so we adopt a backward greedy approximation to find parameters that best explain the latent dynamics. In this approach, we begin with the full set of polynomial terms in Eq~\ref{eq:general_message_function} used during Particle EM, and successively eliminate the least significant polynomial term $G_{abc}$ until the regularized loss stops decreasing.

\subsubsection*{Inferring an example TAP brain}
\label{subsection:methods_infer_example_TAP_brain}
For inferring inference in the example TAP brain described in the previous section, we first generated $2000$ and $25000$ batches of data for the ICA and Particle EM steps, respectively. Each batch of neural activity was generated using inputs with $T=25$ time-steps, 
input correlation time $T_{\textrm{const}} \sim \mathcal{U}\lbr 2, 5\rbr$, and amplitude scaling factor $\tilde{g}_o \sim \mathcal{U}\lbr 5, 25\rbr$. 

For each assumed value of $\hat{N}_s$, we perform ICA to obtain an initial estimate of $\hat{R}$. We also initialize the offset $\hat{\vb}$ to the mean of the neural activity across all batches and neurons. We initialize the remaining parameters as follows. The canonical message parameters $\hat{G}$ were sampled from $\mathcal{N}(0, 0.01)$. For the coupling matrix $\hat{J}$, we generated a dense, symmetric matrix whose elements were sampled from $\mathcal{N}(0, 0.05)$. For the input mapping matrix $\hat{V}$, we used an orthogonal matrix that was obtained from the singular value decomposition of a matrix of size $N_s \times N_o$ with entries sampled from $\mathcal{N}(0,1)$.

For the Particle EM step, we assume that the process noise has a small variance of $10^{-5}$. We also assume prior knowledge of the covariance of the measurement noise. For the amplitude scaling factor settings used to generate our inputs, we set the covariance of the measurement noise to a diagonal matrix with variance of $0.08$ for each neuron. Note that we can also estimate this covariance using neural activity from the TAP brain when the input is held constant for a long duration and the TAP dynamics converge to fixed points for constant inputs. We run Particle EM for $25000$ iterations, using 4 batches of data per iteration. The 4 batches are selected randomly from the $25000$ batches of data at the start of each iteration. For the particle filter, we use $K=100$ particles. For the M-step we use the Adam optimizer with a learning rate of $2 \times 10^{-3}$. We lower this learning rate to $1 \times 10^{-3}$ and $5 \times 10^{-4}$ after $12500$ and $18750$ iterations, respectively. To evaluate the Particle EM estimates, we use $B_{\textrm{test}} = 500$ batches of test data generated using the aforementioned input settings. 

Next, we use the parameters corresponding to the optimal $\hat{N}_s$, to run the particle filter on $2000$ batches of test data (also generated using the same input settings previously described). We used the latent activity obtained from this particle filter to run the greedy backward optimization step to refine the estimates of the coupling matrix and the canonical message parameters. We exclude the terms that are quadratic in both $x_{it}$ and $x_{jt}$ while initializing the greedy search. These terms contribute to one of the dominant degeneracies in our parameterization (see eigenvector 2 in Fig~\ref{fig:Degeneracies}F). For each subset of $\hat{G}$ parameters, we minimize the $\ell_0$-regularized loss function using the Adam optimizer with a learning rate of $10^{-2}$, mini-batch size of $100$, and $4000$ iterations. We use $\alpha = 2 \times 10^{-6}$ as the weighting factor for the $\ell_0$ regularization.


\subsubsection*{MLP based generalization}

To investigate the necessity of the message-passing inductive bias, we explored three relaxations of the setup used in our initial analysis of the TAP brain (Fig~\ref{fig:Example-simulation}). To recap, our initial analysis assumed known functional forms for the aggregation and update functions, and used a polynomial basis (Eq~\ref{eq:general_message_function}) for the message function. Here, we consider the following three MLP based relaxations:
\begin{enumerate}
\item MLP Message Function: This model replaces the polynomial basis for the message function $\mathcal{M}$ with a generic $\text{MLP}$, while retaining the original, a priori known forms for the aggregation (summation) and node update functions. The $\text{MLP}$ is common to all edges and is parameterized as $\text{MLP}(\text{in\_channels}=3, \text{hidden\_channels}=[16, 32, 1], \text{activation\_layer}=\text{torch.nn.GELU})$. The three inputs to the MLP are the state of the receiving node $x_i$, the state of the sending node $x_j$, and the coupling strength $J_{ij}$.
\item MLP Message and MLP Update Functions: This model further relaxes the framework by using an MLP for the message function $\mathcal{M}$ and a second MLP for the node update function $\mathcal{U}$ (which in our canonical model includes the sigmoid nonlinearity). The aggregation function remains fixed as summation.
\begin{itemize}
    \item $\mathcal{M}$: MLP$(\text{in\_channels}=3, \text{hidden\_channels}=[16, 32, 1], \text{activation\_layer}=\text{torch.nn.GELU})$, taking $x_i, x_j, J_{ij}$ as input.
    \item $\mathcal{U}$:  $\text{MLP}(\text{in\_channels}=1, \text{hidden\_channels}=[64, 32, 16, 1], \text{activation\_layer}=\text{torch.nn.GELU})$, taking the aggregated input $\sum_{j} \mathcal{M}_{j\rightarrow i} + (V \vo_t)_i$ as its sole input.
\end{itemize}

\item MLP Full Latent Dynamics: This represents the most substantial relaxation, where a single, large $\text{MLP}$ is used to model the entire latent state transition $\vx_{t+1} = \text{MLP}(\vx_t, \vo_t)$. This variation completely abandons the assumption of an underlying sparse graph structure and the decomposition into message and update components. The $\text{MLP}$ is parameterized as $\text{MLP}(\text{in\_channels}=N_s+N_o,~ \text{hidden\_channels}=[256, 128, N_s], ~\text{activation\_layer}=\text{torch.nn.GELU})$, where $N_s$ is the number of latent variables and $N_o$ is the dimension of the sensory input. The inputs are the full current latent state $\vx_t$ and the sensory observation $\vo_t$.
\end{enumerate}

Unlike the results presented in Fig~\ref{fig:Example-simulation}, the experiments comparing the performance of these variations with the baseline (polynomial basis for the message function) were conducted in a data-constrained regime to highlight the benefit of a correct inductive bias. We restricted the dataset used for both the ICA initialization of the neural embedding matrix $R$ and the subsequent Particle EM training to just 100 batches of data. For Particle EM, we performed 25,000 iterations, using 4 batches of data per iteration (randomly selected from the 100 available batches). The particle filter utilized $K=100$ particles for state estimation. In the M-step, we used the Adam optimizer with a fixed learning rate of $4 \times 10^{-4}$ for all the model variations.

\subsection*{Particle Expectation Maximization}
\label{subsection:particle_EM}

A particle filter approximates the posterior over the latent variables given the current estimate of the model parameters as
\begin{equation}
\hat{p}\lpr \vx_{0:T} | \vr_{0:T};\hat{\vtheta}_n \rpr = \sum_{k=1}^K w^{(k)} \delta \lpr \vx_{0:T} - \vx_{0:T}^{(k)} \rpr
\end{equation}
where $\delta$ is the Dirac delta function. The samples $\lcr \vx_{0:T}^{(k)} \rcr_{k=1}^K$ are referred to as particles. Each particle represents one hypothesis about the state of the system, and the corresponding importance weight $w^{(k)}$ is a measure of how probable that particular latent trajectory is. 
To sample the particle trajectories, we use the following sequential sampling distribution:
\begin{align}
    \pi \lpr \vx_t | \vx_{t-1}, \vr_{t} \rpr &= p \lpr \vx_t | \vx_{t-1}, \vr_{t}; \vtheta_n \rpr \nonumber \\
    &= \frac{p \lpr \vx_t | \vx_{t-1} ; \vtheta_n\rpr  p \lpr \vr_t | \vx_{t} ; \vtheta_n \rpr}{p \lpr \vr_t | \vx_{t-1} ; \vtheta_n \rpr} 
\end{align}
The unnormalized weight for each particle can also be sequentially updated as
\begin{equation}
    \tilde{w}_t^{(k)} \propto \tilde{w}_{t-1}^{(k)} p \lpr \vr_t | \vx_{t-1}^{(k)} ; \vtheta_n \rpr
\end{equation}
Since the conditional distributions in Eqs~\ref{eq:methods-conditional-distributions}a,b are Gaussian, we can analytically derive the sequential sampling distribution and corresponding weight update as:
\begin{subequations}
\begin{align}
p \lpr \vx_t | \vx_{t-1}, \vr_t ; \vtheta_n \rpr &= \mathcal{N}\lpr P^{-1}_n\vv_t, P^{-1}_n \rpr \\
p \lpr \vr_t | \vx_{t-1} ; \vtheta_n \rpr &= Z^{-1} \exp\lcr-\frac{1}{2} \lpr   \boldsymbol{\mu}_{t-1}^{\top}\Sigma_{\xi}^{-1}\boldsymbol{\mu}_{t-1} + \lpr\vr_t-\vb_n\rpr^{\top} \Sigma_{\eta}^{-1}\lpr\vr_t-\vb_n\rpr - \vv_t^{\top}P_n\vv_t \rpr \rcr
\end{align}
\end{subequations}

\noindent where
\begin{subequations}
\begin{align}
    P_n &= \Sigma_{\xi}^{-1} + R_n^{\top} \Sigma_{\eta}^{-1} R_n \\
    \vv_t &= \Sigma_{\xi}^{-1}\boldsymbol{\mu}_{t-1} + R_n^{\top} \Sigma_{\eta}^{-1} \lpr\vr_t-\vb_n\rpr
\end{align}
\end{subequations}
$R_n$ and $\vb_n$ are the current estimates of the neural embedding matrix and offset, and $Z$ is a normalization constant. The conditional mean $\boldsymbol{\mu}_{t-1}$ is obtained using Eq~\ref{eq:methods-conditional-mean} with the current estimates of the input mapping matrix $V_n$, coupling matrix $J_n$, message parameters $G_n$, and inputs $\vo_{t-1}$.

\begin{algorithm}[ht]
\SetAlgoLined
 Initialization\; Given current estimates of parameters $\vtheta_n$,\\ 
 Sample $\vx_0^{(k)} \sim \mathcal{N}\lpr R_n^{\dagger} \lpr\vr_{0}-\vb_n\rpr, \Sigma_{\xi} \rpr$, assign initial weight $w_0^{(k)} = \frac{1}{K}$, for $k=1,...,K$ \\
 Set observed data likelihood $\mathcal{L}_0 = 1$
 
 \For{$t = 1$ $\textrm{to}$ $T$}{
 
    Propagate the trajectories: 
    
    \For{$k = 1$ to $K$}{
    Sample $\vx_t^{(k)} \sim \mathcal{N}\lpr P_n^{-1}\vv^{(k)}_t, P_n^{-1} \rpr$\\
    Update weight, $\tilde{w}_t^{(k)} = w_{t-1}^{(k)}p\lpr \vr_t | \vx_{t-1}^{(k)} ; \vtheta_n \rpr$
    }
    
    Update the likelihood: $\mathcal{L}_t = \mathcal{L}_{t-1}\times \lpr \sum_{k=1}^K \tilde{w}_{t}^{(k)}\rpr$ \\
    Normalize the weights: $w_t^{(k)} = \frac{\tilde{w}_t^{(k)}}{\sum_{j=1}^K \tilde{w}_t^{(j)}}$, $k=1,...,K$
 
  \If{$\hat{K}_{\textrm{eff}} < K/2$}{
   ResampleSystematic$\lcr w_t^{(k)}, \vx_{0:t}^{(k)}\rcr_{k=1}^K$
   }
 }
 \caption{Particle filter}
 \label{Algo:particle-filter}
\end{algorithm}

In practice, the importance weights tend to become degenerate after a few time steps, in the sense that most of the samples have very small weights and thus do not significantly contribute to the approximation of the target distribution. This sample impoverishment problem is circumvented using resampling. If the effective sample size
\begin{equation}
    \hat{K}_{\textrm{eff}} \triangleq \frac{1}{\sum_{k=1}^K \lpr w_t^{(k)}\rpr^2}
\end{equation}
falls below a chosen threshold, the current population of particles are resampled using normalized weights as probabilities of selection. Trajectories with small weights are eliminated, and those with large importance weights are replicated. After resampling, the importance weights are reset to $1/K$. There are several ways of performing resampling; the systematic resampling algorithm \cite{li2015resampling} is used here. 

An important property of the particle filter is that the observed data likelihood can also be updated sequentially as a function of the unnormalized weights $\tilde{w}$ \cite{doucet2011tutorial},
\begin{equation}
    \hat{p}\lpr \vr_{0:t}\rpr = \prod_{\tau=0}^t \lcr \sum_{k=1}^K \tilde{w}_{\tau}^{(k)}\rcr
\end{equation}

\noindent Algorithm \ref{Algo:particle-filter} presents all the steps of the particle filter used in our analysis of the TAP brain models. In each iteration of EM, the approximation to the posterior in terms of the particles can be used to simplify the E-step (Eq~\ref{eq:methods-E-step}) as:

\begin{align}
    Q(\vtheta,\vtheta_n) &= \sum_{k=1}^K w^{(k)}_T \log p \lpr \vx_{0:T}^{(k)}, \vr_{0:T}|\vo_{0:T}; \vtheta \rpr \\
     & = \sum_{k=1}^K w^{(k)}_T \lcr \log p(\vx_0^{(k)}) + \sum_{t=1}^T   \log p \lpr \vx_{t}^{(k)}|\vx_{t-1}^{(k)},\vo_{t-1} ; \vtheta \rpr + \log p \lpr \vr_{t}|\vx_{t}^{(k)} ; \vtheta \rpr \rcr \nonumber
\end{align}

\noindent Ignoring all the constant terms, this can be further simplified as,
\begin{align}
    Q(\vtheta,\vtheta_n) & = -\frac{1}{2}\sum_{k=1}^K w^{(k)}_T\sum_{t=1}^T \Big\{ \lpr \vx_t^{(k)} - \boldsymbol{\mu}_{t-1}^{(k)} \rpr^{\top} \Sigma_{\xi}^{-1}\lpr \vx_t^{(k)} - \boldsymbol{\mu}_{t-1}^{(k)} \rpr \\ \nonumber
    &+ \lpr \vr_t - R\vx_t^{(k)} - \vb \rpr^{\top} \Sigma_{\eta}^{-1}\lpr \vr_t - R\vx_t^{(k)} - \vb \rpr \Big\}
\end{align}
where the conditional mean $\boldsymbol{\mu}_{t-1}^{(k)} = \boldsymbol{\mu}_{t-1}^{(k)} \lpr \vx_{t-1}^{(k)}, \vo_{t-1} ; \vtheta  \rpr$ is obtained from Eq~\ref{eq:methods-conditional-mean}.

The $\mathcal{Q}$ function derived in the E-step is non-convex because the conditional mean is a nonlinear function of the parameters $\vtheta$. Thus, a closed form solution for the maximum is not available. Instead, on each M-step, we perform one iteration of gradient ascent on $Q\lpr \vtheta,\vtheta_n\rpr$ using the Adam optimizer \cite{kingma2015adam}. This optimization does not find the maximum value of $Q\lpr \vtheta,\vtheta_n\rpr$. However, the updated parameters $\vtheta_{n+1}$ satisfy,
\begin{equation}
    Q\lpr \vtheta_{n+1},\vtheta_n\rpr \geq Q\lpr \vtheta_{n},\vtheta_n\rpr.
\end{equation}
This approach, to simply increase and not necessarily maximize $Q\lpr \vtheta,\vtheta_n\rpr$, is known as the Generalized Expectation Maximization (GEM) algorithm and is useful in cases like ours where the maximization in the M-step is difficult \cite{mclachlan2007algorithm}. 

The Particle EM algorithm used in the analysis of the TAP brain is summarized in algorithm \ref{Algo:particle-EM}.

\begin{algorithm}[H]
\SetAlgoLined
 Initialize parameters $\theta_0$\;
 \While{$\textrm{observed~data~likelihood~has~not~converged}$}{
  \begin{enumerate}
      \item Use particle filter (algorithm \ref{Algo:particle-filter}) to obtain $\hat{p}\lpr\vx_{0:T} | \vr_{0:T}, \vo_{0:T} ; \vtheta_n \rpr$ 
      \item E step: use particles $\lcr\vx_{0:T}^{(k)} \rcr_{k=1}^K$ to compute $Q\lpr \vtheta,\vtheta_n\rpr$
      \item M step: obtain $\vtheta_{n+1}$ using one iteration of Adam optimizer on $Q\lpr \vtheta,\vtheta_n\rpr$  
  \end{enumerate}
 }
 \caption{Particle EM}
 \label{Algo:particle-EM}
\end{algorithm}

\section*{Acknowledgments}

The authors thank Andreas Tolias, Kre\v{s}imir Josi\'{c}, and Rich Zemel for helpful conversations.

This work was supported in part NSF CAREER grant 1552868, Air Force Office of Scientific Research (AFOSR) grant FA9550-21-1-0422 in the Cognitive and Computational Neuroscience program, and the Intelligence Advanced Research Projects Activity (IARPA) via Department of Interior/Interior Business Center (DoI/IBC) contract number D16PC00003. The U.S. Government is authorized to reproduce and distribute reprints for Governmental purposes notwithstanding any copyright annotation thereon. Disclaimer: the views and conclusions contained herein are those of the authors and should not be interpreted as necessarily representing the official policies or endorsements, either expressed or implied, of IARPA, DoI/IBC, or the U.S. Government.

\nolinenumbers

%
%

\bibliography{References}

@preamble{ " \newcommand{\noop}[1]{} " }

@techreport{page1999pagerank,
  title={The {PageRank} citation ranking: Bringing order to the web},
  author={Page, Lawrence and Brin, Sergey and Motwani, Rajeev and Winograd, Terry},
  institution={Stanford InfoLab},
  note={Technical Report 1999-66},
  year={1999}
}

@article{murayama2004thouless,
  title={Thouless-Anderson-Palmer approach for lossy compression},
  author={Murayama, Tatsuto},
  journal={Physical Review E},
  volume={69},
  number={3},
  pages={035105},
  year={2004},
  publisher={APS}
}

@article{shamir2000thouless,
  title={Thouless-Anderson-Palmer equations for neural networks},
  author={Shamir, Maoz and Sompolinsky, Haim},
  journal={Physical Review E},
  volume={61},
  number={2},
  pages={1839},
  year={2000},
  publisher={APS}
}

@article{opper2001tractable,
  title={Tractable approximations for probabilistic models: The adaptive Thouless-Anderson-Palmer mean field approach},
  author={Opper, Manfred and Winther, Ole},
  journal={Physical Review Letters},
  volume={86},
  number={17},
  pages={3695},
  year={2001},
  publisher={APS}
}

@article{zhang2024inductive,
  title={Inductive biases of neural network modularity in spatial navigation},
  author={Zhang, Ruiyi and Pitkow, Xaq and Angelaki, Dora E},
  journal={Science Advances},
  volume={10},
  number={29},
  pages={eadk1256},
  year={2024}
}

@article{goyal2022inductive,
  title={Inductive biases for deep learning of higher-level cognition},
  author={Goyal, Anirudh and Bengio, Yoshua},
  journal={Proceedings of the Royal Society A},
  volume={478},
  number={2266},
  pages={20210068},
  year={2022},
  publisher={The Royal Society}
}

@article{gao2017theory,
  title={A theory of multineuronal dimensionality, dynamics and measurement},
  author={Gao, Peiran and Trautmann, Eric and Yu, Byron and Santhanam, Gopal and Ryu, Stephen and Shenoy, Krishna and Ganguli, Surya},
  journal={bioRxiv},
  pages={214262},
  year={2017},
  publisher={Cold Spring Harbor Laboratory}
}

@article{genkin2020moving,
  title={Moving beyond generalization to accurate interpretation of flexible models},
  author={Genkin, Mikhail and Engel, Tatiana A},
  journal={Nature Machine Intelligence},
  volume={2},
  number={11},
  pages={674--683},
  year={2020},
  publisher={Nature Publishing Group UK London}
}

@article{langdon2025latent,
  title={Latent circuit inference from heterogeneous neural responses during cognitive tasks},
  author={Langdon, Christopher and Engel, Tatiana A},
  journal={Nature Neuroscience},
  volume={28},
  pages={665--675},
  year={2025}
}

@article{comon1994independent,
  title={Independent component analysis, a new concept?},
  author={Comon, Pierre},
  journal={Signal Processing},
  volume={36},
  number={3},
  pages={287--314},
  year={1994},
  publisher={Elsevier}
}

@article{vilares2011bayesian,
  title={Bayesian models: the structure of the world, uncertainty, behavior, and the brain},
  author={Vilares, Iris and Kording, Konrad},
  journal={Annals of the New York Academy of Sciences},
  volume={1224},
  number={1},
  pages={22--39},
  year={2011},
  publisher={Wiley Online Library}
}

@inproceedings{lange2022interpolating,
  title={Interpolating between sampling and variational inference with infinite stochastic mixtures},
  author={Lange, Richard D and Benjamin, Ari S and Haefner, Ralf M and Pitkow, Xaq},
  booktitle={Uncertainty in Artificial Intelligence},
  pages={1063--1073},
  year={2022},
  organization={PMLR}
}

@article{starkweather2017dopamine,
  title={Dopamine reward prediction errors reflect hidden-state inference across time},
  author={Starkweather, Clara Kwon and Babayan, Benedicte M and Uchida, Naoshige and Gershman, Samuel J},
  journal={Nature Neuroscience},
  volume={20},
  number={4},
  pages={581--589},
  year={2017},
  publisher={Nature Publishing Group}
}

@article{yu2005uncertainty,
  title={Uncertainty, neuromodulation, and attention},
  author={Yu, Angela J and Dayan, Peter},
  journal={Neuron},
  volume={46},
  number={4},
  pages={681--692},
  year={2005},
  publisher={Elsevier}
}

@article{daw2005uncertainty,
  title={Uncertainty-based competition between prefrontal and dorsolateral striatal systems for behavioral control},
  author={Daw, Nathaniel D and Niv, Yael and Dayan, Peter},
  journal={Nature Neuroscience},
  volume={8},
  number={12},
  pages={1704--1711},
  year={2005},
  publisher={Nature Publishing Group}
}

@article{zeki1974functional,
  title={Functional organization of a visual area in the posterior bank of the superior temporal sulcus of the rhesus monkey},
  author={Zeki, Semir M},
  journal={The Journal of Physiology},
  volume={236},
  number={3},
  pages={549--573},
  year={1974},
  publisher={Wiley Online Library}
}

@article{felleman1991distributed,
  title={Distributed hierarchical processing in the primate cerebral cortex},
  author={Felleman, Daniel J and Van Essen, David C},
  journal={Cerebral Cortex},
  volume={1},
  number={1},
  pages={1--47},
  year={1991}
}

@article{dulberg2022modularity,
  title={Modularity benefits reinforcement learning agents with competing homeostatic drives},
  author={Dulberg, Zack and Dubey, Rachit and Berwian, Isabel M and Cohen, Jonathan D},
  journal={arXiv preprint},
  eprint={2204.06608},
  archivePrefix={arXiv},
  year={2022}
}

@article{yuste2015neuron,
  title={From the neuron doctrine to neural networks},
  author={Yuste, Rafael},
  journal={Nature Reviews Neuroscience},
  volume={16},
  number={8},
  pages={487--497},
  year={2015},
  publisher={Nature Publishing Group}
}

@article{ebitz2021population,
  title={The population doctrine in cognitive neuroscience},
  author={Ebitz, R Becket and Hayden, Benjamin Y},
  journal={Neuron},
  volume={109},
  number={19},
  pages={3055--3068},
  year={2021},
  publisher={Elsevier}
}

@article{ma2022principles,
  title={On the principles of Parsimony and Self-consistency for the emergence of intelligence},
  author={Ma, Yi and Tsao, Doris and Shum, Heung-Yeung},
  journal={Frontiers of Information Technology \& Electronic Engineering},
  volume={23},
  number={9},
  pages={1298--1323},
  year={2022},
}

@article{maass2002real,
  title={Real-time computing without stable states: A new framework for neural computation based on perturbations},
  author={Maass, Wolfgang and Natschl{\"a}ger, Thomas and Markram, Henry},
  journal={Neural Computation},
  volume={14},
  number={11},
  pages={2531--2560},
  year={2002},
  publisher={MIT Press}
}

@article{sinz2019engineering,
  title={Engineering a less artificial intelligence},
  author={Sinz, Fabian H and Pitkow, Xaq and Reimer, Jacob and Bethge, Matthias and Tolias, Andreas S},
  journal={Neuron},
  volume={103},
  number={6},
  pages={967--979},
  year={2019},
  publisher={Elsevier}
}

@article{haefner2016perceptual,
  title={Perceptual decision-making as probabilistic inference by neural sampling},
  author={Haefner, Ralf M and Berkes, Pietro and Fiser, J{\'o}zsef},
  journal={Neuron},
  volume={90},
  number={3},
  pages={649--660},
  year={2016},
  publisher={Elsevier}
}

@article{henaff2020representation,
  title={Representation of visual uncertainty through neural gain variability},
  author={H{\'e}naff, Olivier J and Boundy-Singer, Zoe M and Meding, Kristof and Ziemba, Corey M and Goris, Robbe LT},
  journal={Nature Communications},
  volume={11},
  pages={2513},
  year={2020},
  publisher={Nature Publishing Group}
}

@article{abraham2019active,
  title={Active learning of dynamics for data-driven control using Koopman operators},
  author={Abraham, Ian and Murphey, Todd D},
  journal={IEEE Transactions on Robotics},
  volume={35},
  number={5},
  pages={1071--1083},
  year={2019},
  publisher={IEEE}
}

@INPROCEEDINGS{bruder2019nonlinear,
  author={Bruder, Daniel and Remy, C. David and Vasudevan, Ram},
  booktitle={2019 International Conference on Robotics and Automation (ICRA)}, 
  title={Nonlinear System Identification of Soft Robot Dynamics Using Koopman Operator Theory}, 
  year={2019},
  volume={},
  number={},
  pages={6244-6250},
  }

@book{pearl2018book,
  title={The book of why: the new science of cause and effect},
  author={Pearl, Judea and Mackenzie, Dana},
  year={2018},
  publisher={Basic books}
}

@inproceedings{mi2022connectome,
  title={Connectome-constrained latent variable model of whole-brain neural activity},
  author={Mi, Lu and Xu, Richard and Prakhya, Sridhama and Lin, Albert and Shavit, Nir and Samuel, Aravinthan and Turaga, Srinivas C},
  booktitle={International Conference on Learning Representations},
  year={2022}
}

@book{opper2001advanced,
  title={Advanced Mean Field Methods: Theory and Practice},
  editor={Opper, Manfred and Saad, David},
  year={2001},
  publisher={MIT Press},
  address={Cambridge, MA}
}

@article{glaser2020recurrent,
  title={Recurrent switching dynamical systems models for multiple interacting neural populations},
  author={Glaser, Joshua and Whiteway, Matthew and Cunningham, John P and Paninski, Liam and Linderman, Scott},
  journal={Advances in Neural Information Processing Systems},
  volume={33},
  pages={14867--14878},
  year={2020}
}

@article{musslick2021rationalizing,
  title={Rationalizing constraints on the capacity for cognitive control},
  author={Musslick, Sebastian and Cohen, Jonathan D},
  journal={Trends in Cognitive Sciences},
  volume={25},
  number={9},
  pages={757--775},
  year={2021},
  publisher={Elsevier}
}

@article{logiaco2021thalamic,
  title={Thalamic control of cortical dynamics in a model of flexible motor sequencing},
  author={Logiaco, Laureline and Abbott, LF and Escola, Sean},
  journal={Cell Reports},
  volume={35},
  number={9},
  pages={109090},
  year={2021},
  publisher={Elsevier}
}

@inproceedings{kipf2020contrastive,
  title={Contrastive learning of structured world models},
  author={Kipf, Thomas and van der Pol, Elise and Welling, Max},
  booktitle={International Conference on Learning Representations},
  year={2020}
}

@article{mlynarski2021statistical,
  title={Statistical analysis and optimality of neural systems},
  author={M{\l}ynarski, Wiktor and Hled{\'\i}k, Michal and Sokolowski, Thomas R and Tka{\v{c}}ik, Ga{\v{s}}per},
  journal={Neuron},
  volume={109},
  number={7},
  pages={1227--1241},
  year={2021},
  publisher={Elsevier}
}

@article{chen2019equivalence,
  title={On the equivalence between graph isomorphism testing and function approximation with {GNNs}},
  author={Chen, Zhengdao and Villar, Soledad and Chen, Lei and Bruna, Joan},
  journal={Advances in Neural Information Processing Systems},
  volume={32},
  year={2019}
}

@article{sato2020survey,
  title={A survey on the expressive power of graph neural networks},
  author={Sato, Ryoma},
  journal={arXiv preprint},
  eprint={2003.04078},
  archivePrefix={arXiv},
  year={2020}
}

@book{darwin2004origin,
  title={On the origin of species, 1859},
  author={Darwin, Charles},
  year={2004},
  publisher={Routledge}
}

@article{schmid2010dynamic,
  title={Dynamic mode decomposition of numerical and experimental data},
  author={Schmid, Peter J},
  journal={Journal of Fluid Mechanics},
  volume={656},
  pages={5--28},
  year={2010},
  publisher={Cambridge University Press}
}

@article{kschischang2001factor,
  title={Factor graphs and the sum-product algorithm},
  author={Kschischang, Frank R and Frey, Brendan J and Loeliger, H-A},
  journal={IEEE Transactions on Information Theory},
  volume={47},
  number={2},
  pages={498--519},
  year={2001},
  publisher={IEEE}
}

@article{zhang2020factor,
  title={Factor graph neural networks},
  author={Zhang, Zhen and Wu, Fan and Lee, Wee Sun},
  journal={Advances in Neural Information Processing Systems},
  volume={33},
  pages={8577--8587},
  year={2020}
}

@article{walker2020neural,
  title={A neural basis of probabilistic computation in visual cortex},
  author={Walker, Edgar Y and Cotton, R James and Ma, Wei Ji and Tolias, Andreas S},
  journal={Nature Neuroscience},
  volume={23},
  number={1},
  pages={122--129},
  year={2020},
  publisher={Nature Publishing Group}
}

@article{hoel2013quantifying,
  title={Quantifying causal emergence shows that macro can beat micro},
  author={Hoel, Erik P and Albantakis, Larissa and Tononi, Giulio},
  journal={Proceedings of the National Academy of Sciences},
  volume={110},
  number={49},
  pages={19790--19795},
  year={2013},
  publisher={National Acad Sciences}
}

@article{cumming2016feedforward,
  title={Feedforward and feedback sources of choice probability in neural population responses},
  author={Cumming, Bruce G and Nienborg, Hendrikje},
  journal={Current Opinion in Neurobiology},
  volume={37},
  pages={126--132},
  year={2016},
  publisher={Elsevier}
}

@article{maheswaranathan2019universality,
  title={Universality and individuality in neural dynamics across large populations of recurrent networks},
  author={Maheswaranathan, Niru and Williams, Alex and Golub, Matthew and Ganguli, Surya and Sussillo, David},
  journal={Advances in Neural Information Processing Systems},
  volume={32},
  year={2019}
}

@inproceedings{li2022learning,
      title={Learning Dynamics and Structure of Complex Systems Using Graph Neural Networks},
      author={Li, Zhe and Tolias, Andreas S. and Pitkow, Xaq},
      booktitle={NeurIPS 2022 Workshop: A Causal View on Dynamical Systems},
      year={2022},
      eprint={2202.10996},
      archivePrefix={arXiv},
      primaryClass={cs.LG}
}

@article{wang2005gaussian,
  title={Gaussian process dynamical models},
  author={Wang, Jack and Hertzmann, Aaron and Fleet, David J},
  journal={Advances in Neural Information Processing Systems},
  volume={18},
  year={2005}
}

@InProceedings{qu2019graph,
  title = 	 {{GMNN}: Graph {M}arkov Neural Networks},
  author =       {Qu, Meng and Bengio, Yoshua and Tang, Jian},
  booktitle = 	 {Proceedings of the 36th International Conference on Machine Learning},
  pages = 	 {5241--5250},
  year = 	 {2019},
  editor = 	 {Chaudhuri, Kamalika and Salakhutdinov, Ruslan},
  volume = 	 {97},
  series = 	 {Proceedings of Machine Learning Research},
  publisher =    {PMLR},
}

@article{durstewitz2016computational,
  title={Computational models as statistical tools},
  author={Durstewitz, Daniel and Koppe, Georgia and Toutounji, Hazem},
  journal={Current Opinion in Behavioral Sciences},
  volume={11},
  pages={93--99},
  year={2016},
  publisher={Elsevier}
}

@article{shindo2018inferring,
  title={Inferring a nonlinear biochemical network model from a heterogeneous single-cell time course data},
  author={Shindo, Yuki and Kondo, Yohei and Sako, Yasushi},
  journal={Scientific Reports},
  volume={8},
  pages={6790},
  year={2018},
  publisher={Nature Publishing Group}
}

@article{linderman2017using,
  title={Using computational theory to constrain statistical models of neural data},
  author={Linderman, Scott W and Gershman, Samuel J},
  journal={Current Opinion in Neurobiology},
  volume={46},
  pages={14--24},
  year={2017},
  publisher={Elsevier}
}

@article{jaeger2004harnessing,
author = {Herbert Jaeger  and Harald Haas },
title = {Harnessing Nonlinearity: Predicting Chaotic Systems and Saving Energy in Wireless Communication},
journal = {Science},
volume = {304},
number = {5667},
pages = {78-80},
year = {2004},
}

@article{ghahramani2000variational,
  title={Variational learning for switching state-space models},
  author={Ghahramani, Zoubin and Hinton, Geoffrey E},
  journal={Neural Computation},
  volume={12},
  number={4},
  pages={831--864},
  year={2000},
  publisher={MIT Press}
}

@article{fox2008nonparametric,
  title={Nonparametric Bayesian learning of switching linear dynamical systems},
  author={Fox, Emily and Sudderth, Erik and Jordan, Michael and Willsky, Alan},
  journal={Advances in Neural Information Processing Systems},
  volume={21},
  year={2008}
}

@incollection{Jeb14,
  author={Jebara, Tony},
  title={Perfect graphs and graphical modeling},
  booktitle={Tractability: Practical Approaches to Hard Problems},
  editor={Bordeaux, Lucas and Hamadi, Youssef and Kohli, Pushmeet},
  publisher={Cambridge University Press},
  address={Cambridge},
  pages={39--68},
  year={2014},
}

@inproceedings{ng2000algorithms,
  title={Algorithms for inverse reinforcement learning},
  author={Ng, Andrew Y and Russell, Stuart J},
  booktitle={Proceedings of the Seventeenth International Conference on Machine Learning},
  pages={663--670},
  year={2000}
}

@inproceedings{dvijotham2010inverse,
  title={Inverse optimal control with linearly-solvable {MDPs}},
  author={Dvijotham, Krishnamurthy and Todorov, Emanuel},
  booktitle={Proceedings of the 27th International Conference on Machine Learning},
  pages={335--342},
  year={2010}
}

@inproceedings{jarrett2021inverse,
  title={Inverse decision modeling: Learning interpretable representations of behavior},
  author={Jarrett, Daniel and H{\"u}y{\"u}k, Alihan and {van der Schaar}, Mihaela},
  booktitle={International Conference on Machine Learning},
  pages={4755--4771},
  year={2021},
  organization={PMLR}
}

@article{straub2022putting,
  title={Putting perception into action with inverse optimal control for continuous psychophysics},
  author={Straub, Dominik and Rothkopf, Constantin A},
  journal={eLife},
  volume={11},
  pages={e76635},
  year={2022}
}

@article{chalk2021inferring,
  title={Inferring the function performed by a recurrent neural network},
  author={Chalk, Matthew and Tka{\v{c}}ik, Ga{\v{s}}per and Marre, Olivier},
  journal={PLOS ONE},
  volume={16},
  number={4},
  pages={e0248940},
  year={2021},
  publisher={Public Library of Science San Francisco, CA USA}
}

@inproceedings{sabour2017dynamic,
  title={Dynamic routing between capsules},
  author={Sabour, Sara and Frosst, Nicholas and Hinton, Geoffrey E},
  booktitle={Advances in Neural Information Processing Systems},
  volume={30},
  pages={3856--3866},
  year={2017}
}

@article{xi2017capsule,
  title={Capsule network performance on complex data},
  author={Xi, Edgar and Bing, Selina and Jin, Yang},
  journal={arXiv preprint},
  eprint={1712.03480},
  archivePrefix={arXiv},
  year={2017}
}

@article{schmidt2009distilling,
  title={Distilling free-form natural laws from experimental data},
  author={Schmidt, Michael and Lipson, Hod},
  journal={Science},
  volume={324},
  number={5923},
  pages={81--85},
  year={2009},
  publisher={American Association for the Advancement of Science}
}

@article{orhan2017efficient,
  title={Efficient probabilistic inference in generic neural networks trained with non-probabilistic feedback},
  author={Orhan, A Emin and Ma, Wei Ji},
  journal={Nature Communications},
  volume={8},
  pages={138},
  year={2017},
  publisher={Nature Publishing Group}
}

@article{bakarji2023discovering,
  title={Discovering governing equations from partial measurements with deep delay autoencoders},
  author={Bakarji, Joseph and Champion, Kathleen and Kutz, J Nathan and Brunton, Steven L},
  journal={Proceedings of the Royal Society A},
  volume={479},
  number={2276},
  pages={20230422},
  year={2023},
}

@article{baddoo2023physics,
  title={Physics-informed dynamic mode decomposition},
  author={Baddoo, Peter J and Herrmann, Benjamin and McKeon, Beverley J and Kutz, J Nathan and Brunton, Steven L},
  journal={Proceedings of the Royal Society A},
  volume={479},
  number={2271},
  pages={20220576},
  year={2023},
}

@incollection{takens1981detecting,
  title={Detecting strange attractors in turbulence},
  author={Takens, Floris},
  booktitle={Dynamical Systems and Turbulence, Warwick 1980},
  pages={366--381},
  year={1981},
  publisher={Springer}
}

@article{de2020large,
  title={A large-scale standardized physiological survey reveals functional organization of the mouse visual cortex},
  author={de Vries, Saskia EJ and Lecoq, Jerome A and Buice, Michael A and Groblewski, Peter A and Ocker, Gabriel K and Oliver, Michael and Feng, David and Cain, Nicholas and Ledochowitsch, Peter and Millman, Daniel and others},
  journal={Nature Neuroscience},
  volume={23},
  number={1},
  pages={138--151},
  year={2020},
  publisher={Nature Publishing Group}
}

@article{cisek2019resynthesizing,
  title={Resynthesizing behavior through phylogenetic refinement},
  author={Cisek, Paul},
  journal={Attention, Perception, \& Psychophysics},
  volume={81},
  number={7},
  pages={2265--2287},
  year={2019},
  publisher={Springer}
}

@article{deneve2008bayesian,
  title={Bayesian spiking neurons I: inference},
  author={Deneve, Sophie},
  journal={Neural Computation},
  volume={20},
  number={1},
  pages={91--117},
  year={2008},
  publisher={MIT Press}
}

@article{hyvarinen2000independent,
  title={Independent component analysis: algorithms and applications},
  author={Hyv{\"a}rinen, Aapo and Oja, Erkki},
  journal={Neural Networks},
  volume={13},
  number={4-5},
  pages={411--430},
  year={2000},
  publisher={Elsevier}
}

@article{hyvarinen1999nonlinear,
  title={Nonlinear independent component analysis: Existence and uniqueness results},
  author={Hyv{\"a}rinen, Aapo and Pajunen, Petteri},
  journal={Neural Networks},
  volume={12},
  number={3},
  pages={429--439},
  year={1999},
  publisher={Elsevier}
}

@inproceedings{duncker2019learning,
  title={Learning interpretable continuous-time models of latent stochastic dynamical systems},
  author={Duncker, Lea and Bohner, Gergo and Boussard, Julien and Sahani, Maneesh},
  booktitle={International Conference on Machine Learning},
  pages={1726--1734},
  year={2019},
  organization={PMLR}
}

@article{fei2024higher,
  title={Generalization of graph network inferences in higher-order graphical models},
  author={Fei, Yicheng and Pitkow, Xaq},
  journal={Journal of Applied and Computational Topology},
  volume={8},
  number={5},
  pages={1231--1256},
  year={2024},
}

@article{wolpert1997no,
  title={No free lunch theorems for optimization},
  author={Wolpert, David H and Macready, William G},
  journal={IEEE Transactions on Evolutionary Computation},
  volume={1},
  number={1},
  pages={67--82},
  year={1997},
  publisher={IEEE}
}

@article{zador2019critique,
  title={A critique of pure learning and what artificial neural networks can learn from animal brains},
  author={Zador, Anthony M},
  journal={Nature Communications},
  volume={10},
  pages={3770},
  year={2019},
  publisher={Nature Publishing Group}
}

@book{pearl2009causality,
  title={Causality},
  author={Pearl, Judea},
  year={2009},
  publisher={Cambridge university press}
}

@inproceedings{kipf2018neural,
  title={Neural relational inference for interacting systems},
  author={Kipf, Thomas and Fetaya, Ethan and Wang, Kuan-Chieh and Welling, Max and Zemel, Richard},
  booktitle={Proceedings of the 35th International Conference on Machine Learning},
  series={Proceedings of Machine Learning Research},
  volume={80},
  pages={2688--2697},
  year={2018}
}

@article{turner2017approaches,
  title={Approaches to analysis in model-based cognitive neuroscience},
  author={Turner, Brandon M and Forstmann, Birte U and Love, Bradley C and Palmeri, Thomas J and Van Maanen, Leendert},
  journal={Journal of Mathematical Psychology},
  volume={76},
  pages={65--79},
  year={2017},
  publisher={Elsevier}
}

@article{richards2019deep,
  title={A deep learning framework for neuroscience},
  author={Richards, Blake A and Lillicrap, Timothy P and Beaudoin, Philippe and Bengio, Yoshua and Bogacz, Rafal and Christensen, Amelia and Clopath, Claudia and Costa, Rui Ponte and de Berker, Archy and Ganguli, Surya and Gillon, Colleen J. and Hafner, Danijar and Kepecs, Adam and Kriegeskorte, Nikolaus and Latham, Peter and Lindsay, Grace W and Miller, Kenneth D and Naud, Richard and Pack, Christopher C and Poirazi, Panayiota and Roelfsema, Pieter and Sacramento, João and Saxe, Andrew and Scellier, Benjamin and Schapiro, Anna C and Senn, Walter and Wayne, Greg and Yamins, Daniel and Zenke, Friedemann and Zylberberg, Joel and Therien, Denis and Kording, Konrad P},
  journal={Nature Neuroscience},
  volume={22},
  number={11},
  pages={1761--1770},
  year={2019},
  publisher={Nature Publishing Group}
}

@article{conant1970every,
	title        = {Every good regulator of a system must be a model of that system},
	author       = {Conant, Roger C and Ashby, W Ross},
	year         = 1970,
	journal      = {International Journal of Systems Science},
	publisher    = {Taylor \& Francis},
	volume       = 1,
	number       = 2,
	pages        = {89--97},
}

@inproceedings{goyal2021recurrent,
  title={Recurrent independent mechanisms},
  author={Goyal, Anirudh and Lamb, Alex and Hoffmann, Jordan and Sodhani, Shagun and Levine, Sergey and Bengio, Yoshua and Sch{\"o}lkopf, Bernhard},
  booktitle={International Conference on Learning Representations},
  year={2021},
  eprint={1909.10893},
  archivePrefix={arXiv}
}

@article{friston2010free,
	title        = {The free-energy principle: a unified brain theory?},
	author       = {Friston, Karl},
	year         = 2010,
	journal      = {Nature Reviews Neuroscience},
	publisher    = {Nature publishing group},
	volume       = 11,
	number       = 2,
	pages        = {127--138},
}

@article{wu2020rational,
	title        = {Rational thoughts in neural codes},
	author       = {Wu, Zhengwei and Kwon, Minhae and Daptardar, Saurabh and Schrater, Paul and Pitkow, Xaq},
	year         = 2020,
	journal      = {Proceedings of the National Academy of Sciences},
	publisher    = {National Acad Sciences},
	volume       = 117,
	number       = 47,
	pages        = {29311--29320},
}

@inproceedings{savin2014spatio,
	title        = {Spatio-temporal representations of uncertainty in spiking neural networks},
	author       = {Savin, Cristina and Deneve, Sophie},
	year         = 2014,
	booktitle    = {Advances in Neural Information Processing Systems},
	pages        = {2024--2032},
}

@book{helmholtz1925treatise,
	title        = {Treatise on Physiological Optics},
	volume       = {III},
	author       = {Helmholtz, Hermann},
	note         = {Translated from the 3rd German edition by J. P. C. Southall},
	publisher    = {Optical Society of America},
	address      = {Menasha, WI},
	year         = 1925,
}

@article{churchland2012neural,
	title        = {Neural population dynamics during reaching},
	author       = {Churchland, Mark M and Cunningham, John P and Kaufman, Matthew T and Foster, Justin D and Nuyujukian, Paul and Ryu, Stephen I and Shenoy, Krishna V},
	year         = 2012,
	journal      = {Nature},
	publisher    = {Nature Publishing Group},
	volume       = 487,
	number       = 7405,
  pages={51--56},
}

@inproceedings{kwon2020inverse,
	title        = {Inverse rational control with partially observable continuous nonlinear dynamics},
	author       = {Kwon, Minhae and Daptardar, Saurabh and Schrater, Paul and Pitkow, Xaq},
	booktitle    = {Advances in Neural Information Processing Systems},
	volume       = {33},
	year         = {2020},
	eprint       = {1908.04696},
	archivePrefix = {arXiv}
}

@article{koopman1931hamiltonian,
	title        = {Hamiltonian systems and transformation in Hilbert space},
	author       = {Koopman, Bernard O},
	year         = 1931,
	journal      = {Proceedings of the National Academy of Sciences},
	publisher    = {National Academy of Sciences},
	volume       = 17,
	number       = 5,
  pages={315--318},
}

@article{pandarinath2018inferring,
	title        = {Inferring single-trial neural population dynamics using sequential auto-encoders},
	author       = {Pandarinath, Chethan and O’Shea, Daniel J and Collins, Jasmine and Jozefowicz, Rafal and Stavisky, Sergey D and Kao, Jonathan C and Trautmann, Eric M and Kaufman, Matthew T and Ryu, Stephen I and Hochberg, Leigh R and others},
	year         = 2018,
	journal      = {Nature Methods},
	publisher    = {Nature Publishing Group},
  pages={805--815},
  volume={15},
  number={10}
}

@article{giahi2019dynamic,
  title={Dynamic brain interactions during picture naming},
  author={Giahi Saravani, Aram and Forseth, Kiefer J and Tandon, Nitin and Pitkow, Xaq},
  journal={eNeuro},
  volume={6},
  number={4},
  pages={ENEURO.0472-18.2019},
  year={2019}
}

@inproceedings{babes2011apprenticeship,
	title        = {Apprenticeship learning about multiple intentions},
	author       = {Babes, Monica and Marivate, Vukosi and Subramanian, Kaushik and Littman, Michael L},
	year         = 2011,
	booktitle    = {Proceedings of the 28th International Conference on Machine Learning (ICML-11)},
	pages        = {897--904},
}

@inproceedings{schmitt2017see,
	title        = {I See What You See: Inferring Sensor and Policy Models of Human Real-World Motor Behavior},
	author       = {Schmitt, Felix and Bieg, Hans-Joachim and Herman, Michael and Rothkopf, Constantin A},
	year         = 2017,
	booktitle    = {Proceedings of the AAAI Conference on Artificial Intelligence},
	pages        = {3797--3803},
}

@article{musall2019single,
  title={Single-trial neural dynamics are dominated by richly varied movements},
  author={Musall, Simon and Kaufman, Matthew T and Juavinett, Ashley L and Gluf, Steven and Churchland, Anne K},
  journal={Nature Neuroscience},
  volume={22},
  number={10},
  pages={1677--1686},
  year={2019}
}

@article{yang2021revealing,
	title        = {Revealing nonlinear neural decoding by analyzing choices},
	author       = {Yang, Qianli and Walker, Edgar Y and Cotton, R James and Tolias, Andreas Savas and Pitkow, Xaq},
	journal      = {Nature Communications},
	volume       = {12},
	year         = {2021},
  pages={6557}
}

@article{whiteway2019quest,
	title        = {The quest for interpretable models of neural population activity},
	author       = {Whiteway, Matthew R and Butts, Daniel A},
	year         = 2019,
	journal      = {Current Opinion in Neurobiology},
	publisher    = {Elsevier},
	volume       = 58,
	pages        = {86--93},
}

@inproceedings{vertes2018flexible,
	title        = {Flexible and accurate inference and learning for deep generative models},
	author       = {V{\'e}rtes, Eszter and Sahani, Maneesh},
	year         = 2018,
	booktitle    = {Advances in Neural Information Processing Systems},
	pages        = {4166--4175},
}

@article{glascher2010states,
	title        = {States versus rewards: dissociable neural prediction error signals underlying model-based and model-free reinforcement learning},
	author       = {Gl{\"a}scher, Jan and Daw, Nathaniel and Dayan, Peter and O'Doherty, John P},
	year         = 2010,
	journal      = {Neuron},
	publisher    = {Elsevier},
	volume       = 66,
	number       = 4,
	pages        = {585--595},
}

@article{turner2013bayesian,
	title        = {A Bayesian framework for simultaneously modeling neural and behavioral data},
	author       = {Turner, Brandon M and Forstmann, Birte U and Wagenmakers, Eric-Jan and Brown, Scott D and Sederberg, Per B and Steyvers, Mark},
	year         = 2013,
	journal      = {NeuroImage},
	publisher    = {Elsevier},
	volume       = 72,
	pages        = {193--206},
}

@article{pitkow2015can,
	title        = {How can single sensory neurons predict behavior?},
	author       = {Pitkow, Xaq and Liu, Sheng and Angelaki, Dora E and DeAngelis, Gregory C and Pouget, Alexandre},
	year         = 2015,
	journal      = {Neuron},
	publisher    = {Elsevier},
	volume       = 87,
	number       = 2,
	pages        = {411--423},
}

@article{ackley1985learning,
  title={A learning algorithm for {Boltzmann} machines},
	author       = {Ackley, David H and Hinton, Geoffrey E and Sejnowski, Terrence J},
	year         = 1985,
	journal      = {Cognitive Science},
	publisher    = {Wiley Online Library},
	volume       = 9,
	number       = 1,
	pages        = {147--169},
}

@article{georgopoulos1986neuronal,
	title        = {Neuronal population coding of movement direction},
	author       = {Georgopoulos, Apostolos P and Schwartz, Andrew B and Kettner, Ronald E},
	year         = 1986,
	journal      = {Science},
	publisher    = {American Association for the Advancement of Science},
	volume       = 233,
	number       = 4771,
	pages        = {1416--1419},
}

@article{berkes2011sampling,
	title        = {Spontaneous cortical activity reveals hallmarks of an optimal internal model of the environment},
	author       = {Berkes, P. and Orb{\'a}n, G. and Lengyel, M. and Fiser, J.},
	year         = 2011,
	journal      = {Science},
	volume       = 331,
	number       = 6013,
	pages        = {83--7},
}

@article{hubel1962receptive,
	title        = {Receptive fields, binocular interaction and functional architecture in the cat's visual cortex},
	author       = {Hubel, David H and Wiesel, Torsten N},
	year         = 1962,
	journal      = {The Journal of Physiology},
	publisher    = {Wiley Online Library},
	volume       = 160,
	number       = 1,
	pages        = {106--154},
}

@article{thouless1977,
  title={Solution of `{Solvable} model of a spin glass'},
  author={Thouless, David J and Anderson, Philip W and Palmer, Robert G},
  journal={Philosophical Magazine},
  volume={35},
  number={3},
  pages={593--601},
  year={1977},
  publisher={Taylor \& Francis}
}

@inproceedings{raju2016inference,
	title={Inference by reparameterization in neural population codes},
	author={Raju, Rajkumar Vasudeva and Pitkow, Xaq},
	booktitle={Advances in Neural Information Processing Systems},
	pages={2029--2037},
	year={2016}
}

@article{scarselli2009graph,
	title={The graph neural network model},
	author={Scarselli, Franco and Gori, Marco and Tsoi, Ah Chung and Hagenbuchner, Markus and Monfardini, Gabriele},
	journal={IEEE Transactions on Neural Networks},
	volume={20},
	number={1},
	pages={61--80},
	year={2009},
	publisher={IEEE}
}

@article{battaglia2018relational,
  title={Relational inductive biases, deep learning, and graph networks},
  author={Battaglia, Peter W and Hamrick, Jessica B and Bapst, Victor and Sanchez-Gonzalez, Alvaro and Zambaldi, Vinicius and Malinowski, Mateusz and Tacchetti, Andrea and Raposo, David and Santoro, Adam and Faulkner, Ryan and others},
  journal={arXiv preprint},
  eprint={1806.01261},
  archivePrefix={arXiv},
  year={2018}
}

@article{archer2015,
   author = {Archer, Evan and Park, Il Memming and Buesing, Lars and Cunningham, John and Paninski, Liam},
   title = {Black box variational inference for state space models},
   journal = {arXiv preprint},
   eprint = {1511.07367},
   archivePrefix = {arXiv},
   year = {2015}
}

@article{carandini2012normalization,
  title={Normalization as a canonical neural computation},
  author={Carandini, Matteo and Heeger, David J},
  journal={Nature Reviews Neuroscience},
  volume={13},
  number={1},
  pages={51--62},
  year={2012},
  publisher={Nature Publishing Group}
}

@book{pearl1988probabilistic,
  title={Probabilistic reasoning in intelligent systems: networks of plausible inference},
  author={Pearl, Judea},
  year={1988},
  publisher={Morgan Kaufmann}
}

@article{lee2003hierarchical,
  title={Hierarchical Bayesian inference in the visual cortex},
  author={Lee, Tai Sing and Mumford, David},
  journal={Journal of the Optical Society of America A},
  volume={20},
  number={7},
  pages={1434--1448},
  year={2003},
  publisher={Optical Society of America}
}

@book{doya2007bayesian,
  title={Bayesian Brain: Probabilistic Approaches to Neural Coding},
  editor={Doya, Kenji and Ishii, Shin and Pouget, Alexandre and Rao, Rajesh P N},
  year={2007},
  publisher={MIT Press},
  address={Cambridge, MA}
}

@article{rigotti2013importance,
  title={The importance of mixed selectivity in complex cognitive tasks},
  author={Rigotti, Mattia and Barak, Omri and Warden, Melissa R and Wang, Xiao-Jing and Daw, Nathaniel D and Miller, Earl K and Fusi, Stefano},
  journal={Nature},
  volume={497},
  number={7451},
  pages={585--590},
  year={2013},
  publisher={Nature Publishing Group}
}

@article{ma2006bayesian,
  title={Bayesian inference with probabilistic population codes},
  author={Ma, Wei Ji and Beck, Jeffrey M and Latham, Peter E and Pouget, Alexandre},
  journal={Nature Neuroscience},
  volume={9},
  number={11},
  pages={1432--1438},
  year={2006},
  publisher={Nature Publishing Group}
}

@article{haefner2013inferring,
  title={Inferring decoding strategies from choice probabilities in the presence of correlated variability},
  author={Haefner, Ralf M and Gerwinn, Sebastian and Macke, Jakob H and Bethge, Matthias},
  journal={Nature Neuroscience},
  volume={16},
  number={2},
  pages={235--242},
  year={2013},
  publisher={Nature Publishing Group}
}

@article{moreno:2014information,
  title={Information-limiting correlations},
  author={Moreno-Bote, Rub{\'e}n and Beck, Jeffrey and Kanitscheider, Ingmar and Pitkow, Xaq and Latham, Peter and Pouget, Alexandre},
  journal={Nature Neuroscience},
  year={2014},
  volume={17},
  pages={1410--1417},
  publisher={Nature Publishing Group}
}

@article{cappe2007overview,
	title={An overview of existing methods and recent advances in sequential Monte Carlo},
	author={Capp{\'e}, Olivier and Godsill, Simon J and Moulines, Eric},
	journal={Proceedings of the IEEE},
	volume={95},
	number={5},
	pages={899--924},
	year={2007},
	publisher={IEEE}
}

@article{kantas2015particle,
	title={On particle methods for parameter estimation in state-space models},
	author={Kantas, Nikolas and Doucet, Arnaud and Singh, Sumeetpal S and Maciejowski, Jan and Chopin, Nicolas},
	journal={Statistical Science},
	volume={30},
	number={3},
	pages={328--351},
	year={2015},
	publisher={Institute of Mathematical Statistics}
}

@article{saxena2019towards,
	title={Towards the neural population doctrine},
	author={Saxena, Shreya and Cunningham, John P},
	journal={Current Opinion in Neurobiology},
	volume={55},
	pages={103--111},
	year={2019},
	publisher={Elsevier}
}

@article{AtomsofNeuralComputation,
	title	= {The atoms of neural computation},
	author	= {Gary Marcus and Adam Marblestone and Tom Dean},
	year	= {2014},
	journal	= {Science},
  pages={551--552},
	volume	= {346},
  number={6209}
}

@book{marr1982vision,
	title={Vision: A computational investigation into the human representation and processing of visual information},
	author={Marr, David},
	year={1982},
	publisher={W. H. Freeman},
	address={San Francisco}
}

@inproceedings{minka2001expectation,
	title={Expectation propagation for approximate Bayesian inference},
	author={Minka, Thomas P},
	booktitle={Proceedings of the Seventeenth Conference on Uncertainty in Artificial Intelligence},
	pages={362--369},
	year={2001},
	organization={Morgan Kaufmann Publishers Inc.}
}

@inproceedings{kingma2015adam,
  title={Adam: A method for stochastic optimization},
  author={Kingma, Diederik P and Ba, Jimmy},
  booktitle={International Conference on Learning Representations},
  year={2015},
  eprint={1412.6980},
  archivePrefix={arXiv}
}

@article{miller2016canonical,
  title={Canonical computations of cerebral cortex},
  author={Miller, Kenneth D},
  journal={Current Opinion in Neurobiology},
  volume={37},
  pages={75--84},
  year={2016},
  publisher={Elsevier}
}

@inproceedings{hoyer2003interpreting,
  title={Interpreting neural response variability as Monte Carlo sampling of the posterior},
  author={Hoyer, Patrik O and Hyv{\"a}rinen, Aapo},
  booktitle={Advances in Neural Information Processing Systems},
  pages={293--300},
  year={2003}
}

@article{williamson2019bridging,
  title={Bridging large-scale neuronal recordings and large-scale network models using dimensionality reduction},
  author={Williamson, Ryan C and Doiron, Brent and Smith, Matthew A and Yu, Byron M},
  journal={Current Opinion in Neurobiology},
  volume={55},
  pages={40--47},
  year={2019},
  publisher={Elsevier}
}

@article{cunningham2014dimensionality,
  title={Dimensionality reduction for large-scale neural recordings},
  author={Cunningham, John P and Yu, Byron M},
  journal={Nature Neuroscience},
  volume={17},
  number={11},
  pages={1500--1509},
  year={2014},
  publisher={Nature Publishing Group}
}

@article{brunton2016discovering,
  title={Discovering governing equations from data by sparse identification of nonlinear dynamical systems},
  author={Brunton, Steven L and Proctor, Joshua L and Kutz, J Nathan},
  journal={Proceedings of the National Academy of Sciences},
  volume={113},
  number={15},
  pages={3932--3937},
  year={2016},
  publisher={National Acad Sciences}
}

@article{mante2013context,
  title={Context-dependent computation by recurrent dynamics in prefrontal cortex},
  author={Mante, Valerio and Sussillo, David and Shenoy, Krishna V and Newsome, William T},
  journal={Nature},
  volume={503},
  number={7474},
  pages={78--84},
  year={2013},
  publisher={Nature Publishing Group}
}

@inproceedings{gao2016linear,
  title={Linear dynamical neural population models through nonlinear embeddings},
  author={Gao, Yuanjun and Archer, Evan W and Paninski, Liam and Cunningham, John P},
  booktitle={Advances in Neural Information Processing Systems},
  pages={163--171},
  year={2016}
}

@inproceedings{linderman2017bayesian,
  title={Bayesian learning and inference in recurrent switching linear dynamical systems},
  author={Linderman, Scott and Johnson, Matthew and Miller, Andrew and Adams, Ryan and Blei, David and Paninski, Liam},
  booktitle={Artificial Intelligence and Statistics},
  pages={914--922},
  year={2017}
}

@article{dempster1977EM,
  title={Maximum likelihood from incomplete data via the EM algorithm},
  author={Dempster, Arthur P and Laird, Nan M and Rubin, Donald B},
  journal={Journal of the Royal Statistical Society: Series B (Methodological)},
  volume={39},
  number={1},
  pages={1--22},
  year={1977},
  publisher={Wiley Online Library}
}

@incollection{doucet2011tutorial,
  title={A tutorial on particle filtering and smoothing: Fifteen years later},
  author={Doucet, Arnaud and Johansen, Adam M},
  booktitle={The Oxford Handbook of Nonlinear Filtering},
  editor={Crisan, Dan and Rozovski{\u{\i}}, Boris},
  publisher={Oxford University Press},
  address={Oxford},
  pages={656--704},
  year={2011}
}

@article{li2015resampling,
  title={Resampling methods for particle filtering: classification, implementation, and strategies},
  author={Li, Tiancheng and Bolic, Miodrag and Djuric, Petar M},
  journal={IEEE Signal Processing Magazine},
  volume={32},
  number={3},
  pages={70--86},
  year={2015},
  publisher={IEEE}
}

@book{mclachlan2007algorithm,
  title={The EM algorithm and extensions},
  author={McLachlan, Geoffrey and Krishnan, Thriyambakam},
  year={2007},
  publisher={John Wiley \& Sons}
}

@inproceedings{yoon2018inference,
  title={Inference in probabilistic graphical models by graph neural networks},
  author={Yoon, KiJung and Liao, Renjie and Xiong, Yuwen and Zhang, Lisa and Fetaya, Ethan and Urtasun, Raquel and Zemel, Richard and Pitkow, Xaq},
  booktitle={International Conference on Learning Representations, Workshop Track},
  year={2018},
  eprint={1803.07710},
  archivePrefix={arXiv}
}

@article{pitkow2017inference,
  title={Inference in the brain: statistics flowing in redundant population codes},
  author={Pitkow, Xaq and Angelaki, Dora E},
  journal={Neuron},
  volume={94},
  number={5},
  pages={943--953},
  year={2017},
  publisher={Elsevier}
}

@book{koller2009probabilistic,
  title={Probabilistic graphical models: principles and techniques},
  author={Koller, Daphne and Friedman, Nir},
  year={2009},
  publisher={MIT press}
}

@article{mardia2008multivariate,
  title={A multivariate von Mises distribution with applications to bioinformatics},
  author={Mardia, Kanti V and Hughes, Gareth and Taylor, Charles C and Singh, Harshinder},
  journal={Canadian Journal of Statistics},
  volume={36},
  number={1},
  pages={99--109},
  year={2008},
  publisher={Wiley Online Library}
}

@article{semedo_cortical_2019,
  title={Cortical areas interact through a communication subspace},
    volume = {102},
    number = {1},
    journal = {Neuron},
    author = {Semedo, João D. and Zandvakili, Amin and Machens, Christian K. and Yu, Byron M. and Kohn, Adam},
    year = {2019},
    pages = {249--259.e4},
}

@article{allier_graph_2026,
    title = {Graph neural networks uncover structure and functions underlying the activity of simulated neural assemblies},
    author = {Allier, Cédric and Heinrich, Larissa and Schneider, Magdalena and Saalfeld, Stephan},
    journal = {arXiv preprint},
  eprint={2602.13325},
  archivePrefix={arXiv},
    year = {2026},
}

@article{bae_functional_2025,
    title = {Functional connectomics spanning multiple areas of mouse visual cortex},
    volume = {640},
    number = {8058},
    journal = {Nature},
    author = {{MICrONS Consortium}},
    year = {2025},
    pages = {435--447},
}

@article{wainwright2008graphical,
  title={Graphical models, exponential families, and variational inference},
  author={Wainwright, Martin J and Jordan, Michael I},
  journal={Foundations and Trends in Machine Learning},
  volume={1},
  number={1--2},
  pages={1--305},
  year={2008}
}

@article{brunton2017chaos,
  title={Chaos as an intermittently forced linear system},
  author={Brunton, Steven L and Brunton, Bingni W and Proctor, Joshua L and Kaiser, Eurika and Kutz, J Nathan},
  journal={Nature Communications},
  volume={8},
  pages={19},
  year={2017}
}

@article{mountcastle1997columnar,
  title={The columnar organization of the neocortex},
  author={Mountcastle, Vernon B},
  journal={Brain},
  volume={120},
  number={4},
  pages={701--722},
  year={1997}
}

\end{document}